 \documentclass[final,5p,times,twocolumn,authoryear,numafflabel]{elsarticle}

\usepackage{titlesec}
\usepackage{placeins}
\usepackage{doi}

\titleformat{\paragraph}
{\normalfont\normalsize\itshape}{\theparagraph}{1em}{}
\titlespacing*{\paragraph}
{0pt}{3.25ex plus 1ex minus .2ex}{1.5ex plus .2ex}

\usepackage{amssymb}
\usepackage{lipsum}
\usepackage{subcaption}
\usepackage{float}
\usepackage{amsmath}
\usepackage{dblfloatfix}
\usepackage{comment}
\usepackage{colortbl}
\usepackage{xcolor}
\usepackage{hyperref} 
\usepackage{multirow}
\usepackage{multicol}
\usepackage{adjustbox}
\usepackage{array}
\usepackage{amssymb}% http://ctan.org/pkg/amssymb
\usepackage{pifont}% http://ctan.org/pkg/pifont
\usepackage{tabularx}
\newcommand{\cmark}{\ding{51}}% check
\newcommand{\xmark}{\ding{55}}% cross

\journal{Ecological Informatics}

\begin{document}

\definecolor{blue}{RGB}{20, 105, 181}
\definecolor{green}{RGB}{104, 200, 59}
\definecolor{orange}{RGB}{249, 102, 44}

\newcommand{\ROne}{\textcolor{blue}{R1}\@\xspace}
\newcommand{\RTwo}{\textcolor{green}{R2}\@\xspace}
\newcommand{\RThree}{\textcolor{orange}{R3}\@\xspace}
\newcommand{\todo}[1]{\textcolor{red}{\textbf{TODO:} #1}}

\newcommand{\todoo}[1]{\textcolor{orange}{\textbf{TODO:} #1}}

\begin{frontmatter}



% \title{Decoding the Sounds of Doñana: Advancements in Bird Detection and Identification Through Deep Learning}
\title{A Bird Song Detector for improving bird identification through Deep Learning: a case study from Doñana}


\author[1,2,3]{Alba Márquez-Rodríguez\textsuperscript{*}}
\author[2,4]{Miguel Ángel Mohedano-Munoz}
\author[1]{Manuel J. Mar\'in-Jim\'enez}
\author[2]{Eduardo Santamaría-García}
\author[5]{Giulia Bastianelli}
\author[2,6]{Pedro Jordano}
\author[2,6]{Irene Mendoza}

\affiliation[1]{organization={University of Córdoba},%Department and Organization
            department={Dept. of Computing and Numeric Analysis,},
            %addressline={}, 
            city={Córdoba},
            %postcode={}, 
            %state={Córdoba},
            country=Spain}
\affiliation[2]{organization={Estación Biológica de Doñana},
            department={Dept. of Ecology and Evolution,},
            %addressline={}, 
            city={Sevilla},
            %postcode={}, 
             %state={Sevilla},
            country=Spain}
\affiliation[3]{organization={University of Cádiz},
            department={Instituto Universitario de Investigación Marina (INMAR), Campus de Excelencia Internacional del Mar (CEIMAR),},
            %addressline={}, 
            city={Puerto Real},
            %postcode={}, 
             state={Cádiz},
            country=Spain}
\affiliation[4]{
            organization={Universidad Politécnica de Madrid},
            department={Escuela Técnica Superior de Ingeniería, Agronómica, Alimentaria y de Biosistemas,},
            %addressline={}, 
            city={Madrid},
            %postcode={}, 
             %state={Sevilla},
            country=Spain}
\affiliation[5]{
            organization={Estación Biológica de Doñana},
            department={ICTS-Doñana (Infraestructura Científico-Técnica Singular de Doñana),},
            %addressline={}, 
            city={Sevilla},
            %postcode={}, 
             %state={Sevilla},
            country=Spain}
\affiliation[6]{
            organization={University of Sevilla},
            department={Dep. of Plant Biology and Ecology,},
            %addressline={}, 
            city={Sevilla},
            %postcode={}, 
             %state={Sevilla},
            country=Spain}

\cortext[cor1]{Corresponding author: \\ Email: alba.marquez@uca.es (Alba Márquez-Rodríguez)}

\begin{abstract}

Passive Acoustic Monitoring (PAM), which uses devices like automatic audio recorders, has become a fundamental tool in conserving and managing natural ecosystems. However, the large volume of unsupervised audio data that PAM generates poses a major challenge for extracting meaningful information. Deep Learning techniques, particularly automated species identification models based on computer vision, offer a promising solution. BirdNET, a widely used model for bird identification, has shown success in many study systems but is limited at local scale due to biases in its training data, which focus on specific locations and target sounds rather than entire soundscapes. A key challenge in bird species detection is that many recordings either lack target species or contain overlapping vocalizations, complicating automatic identification. To overcome these problems, we developed a three-stage pipeline for automatic bird vocalization identification in Doñana National Park (SW Spain), a wetland facing significant conservation threats. We deployed AudioMoth recorders in three main habitats across nine different locations within Doñana, and the manual annotation of 461 minutes of audio data, resulting in 3749 annotations covering 34 classes. Our working pipeline included, first, the development of a Bird Song Detector to isolate bird vocalizations, using spectrograms as graphical representations of bird audio data and applying image processing methods. Second, we classified bird species training custom classifiers at the local scale with BirdNET's embeddings. The best-performing detection model incorporated synthetic background audios through data augmentation and an environmental sound library (ESC-50). Applying the Bird Song Detector before classification improved species identification, as all classification models performed better when analyzing only the segments where birds were detected. Specifically, the combination of the Bird Song Detector and fine-tuned BirdNET increased weighted precision (from 0.18 to 0.37), recall (from 0.21 to 0.30), and F1 score (from 0.17 to 0.28), compared to the baseline without the Bird Song Detector. Our approach demonstrated the effectiveness of integrating a Bird Song Detector with fine-tuned classification models for bird identification at local soundscapes. These findings highlight the need to adapt general-purpose tools for specific ecological challenges, as demonstrated in Doñana. Automatically detecting bird species serves for tracking the health status of this threatened ecosystem, given the sensitivity of birds to environmental changes, and helps in the design of conservation measures for reducing biodiversity loss. 

\end{abstract}



\begin{keyword}
%% keywords here, in the form: keyword \sep keyword, up to a maximum of 6 keywords
AudioMoth \sep Computer Vision \sep Convolutional Neural Networks \sep Ecoacoustics \sep Ecosystem health \sep Passive Acoustic Monitoring %\sep BIRDeep 



\end{keyword}

\end{frontmatter}

%\tableofcontents


\section{Introduction}
\label{introduction}

% State the objectives of the work and provide an adequate background, avoiding a detailed literature survey or a summary of the results.

Natural environments face significant challenges in terms of conservation due to different drivers of global change, such as habitat loss, climate change, species invasion or anthropogenic pressure. In response to this crisis, biodiversity monitoring and species interaction assessments have become essential to understanding environmental impacts and developing conservation strategies. Effective biodiversity monitoring is fundamental for conservation efforts, as it provides the data necessary to make informed decisions. Still, it is challenging to get the necessary data at large spatio-temporal scales.

In this regard, identifying and tracking the presence of birds is crucial, as birds serve as indicators of ecosystem health \citep{gregory2010wild}. Although various automatic monitoring technologies, such as cameras and audio recorders, are already in use, efficiently managing and analyzing the large volumes of data generated by these devices remains a challenge. An effective technology is Passive Acoustic Monitoring (PAM), which uses audio recorders to continuously capture sounds of an environment. PAM is particularly valuable for monitoring biodiversity, as it can operate in remote and inaccessible areas, providing continuous data without disturbing habitat \citep{sugai2019terrestrial}. Using PAM, the spatiotemporal scale of monitoring can be significantly expanded, allowing more comprehensive and detailed ecological studies \citep{gibb2019emerging}. 

In recent years, the cost of automatic recording devices has been reduced, leading to an increase in the collection of this type of data for ecological studies \citep{farley2018situating, darras2019autonomous, metcalf2023good}. Despite their advantages, they generate large amounts of unlabeled data, making analysis difficult and limiting their utility for decision making \citep{tuia2022perspectives}. The primary objective of any PAM project is to address the challenge of efficiently extracting biodiversity information from large volumes of audio data. Thanks to this process, the identification of bird species is automated, which greatly improves data analysis efficiency in the context of environmental monitoring.

Historically, early bird vocalization recognition methods relied on basic sound feature analysis, using techniques such as Random Forests for classification. These methods focused on extracting specific audio features -- such as frequency, pitch, or duration -- to create a feature set that could be used for classification \citep{keen2021machine}. Despite of being effective to a certain extent, these approaches were limited by their reliance on manually crafted features and often struggled with complex and overlapping sounds. Such challenges in traditional feature engineering approaches are well-documented in foundational works on Machine Learning \citep{hastie2009elements,murphy2012machine}, where the need for automated, high-level feature extraction was emphasized as a challenge for accurate classification in complex domains. The recent shift towards Deep Learning \citep{goodfellow2016deep}, has enabled more advanced architectures that autonomously learn rich representations from data, addressing many limitations of earlier methods.

\begin{figure*}[!ht]
    \centering
    \includegraphics[width=1\linewidth]{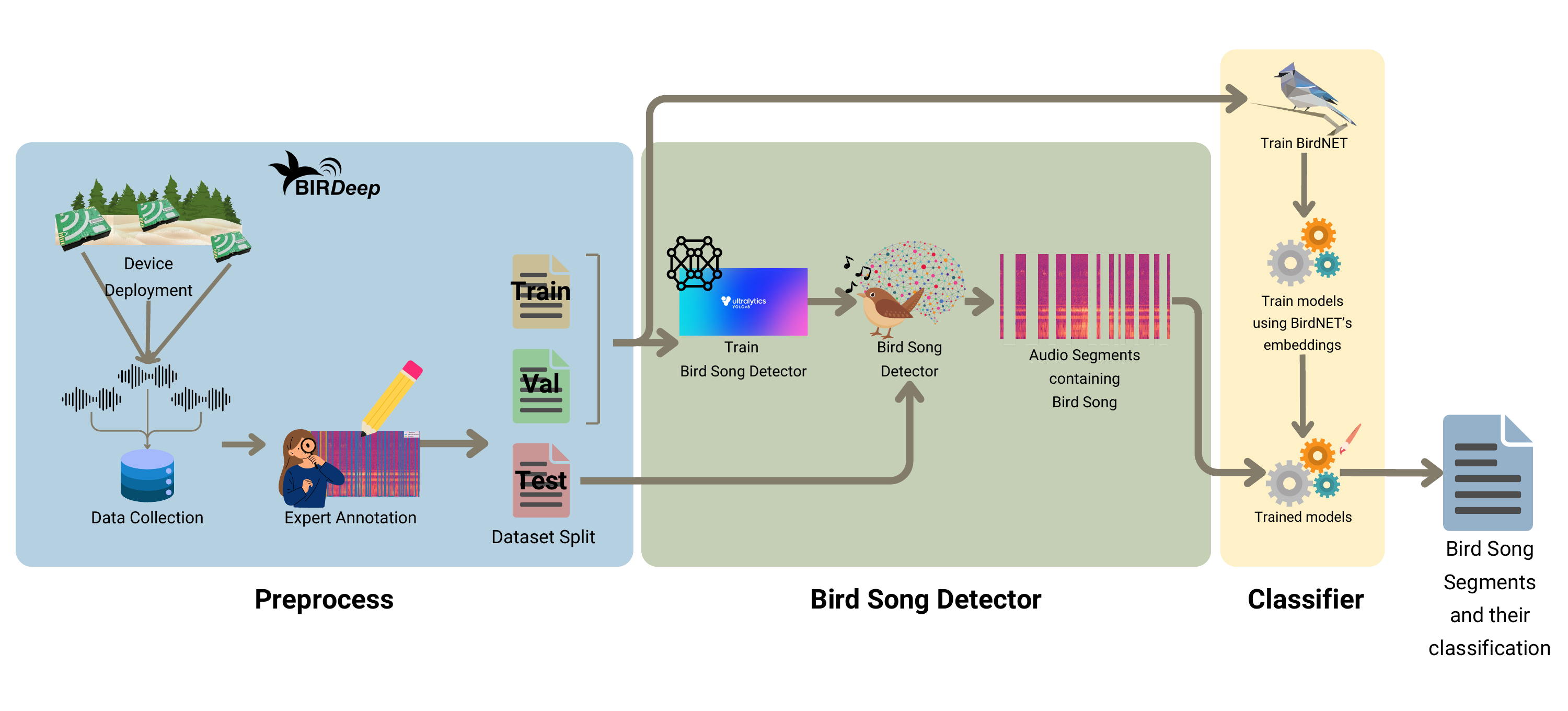}
    \caption{Pipeline used for the development of our Bird Song Detector. The process was divided into three main stages: 
    (1) \textit{Preprocess}: This stage involved deploying automatic recording devices (AudioMoths) in natural habitats of Doñana to collect audio data, as a part of the BIRDeep project. Audio recordings were then annotated by experts to identify bird vocalizations, followed by splitting the dataset into training, validation, and test sets. 
    (2) \textit{Bird Song Detector}: We trained the Bird Song Detector using the annotated dataset. Our Detector was derived from YOLOv8, which is a state-of-the-art model suitable for detecting objects in images. The Bird Song Detector was developed to identify segments of the audio recordings that contained the sonogram of a bird vocalization (only the presence/absence of any bird species). After training, the Detector model was applied to the test dataset, producing segments that contained potential bird vocalizations. 
    (3) \textit{Classifier}: The final stage consisted first of a fine-tuning of BirdNET model, using the audio from Doñana annotated by experts. Second, this fine-tuned BirdNET model was then used to extract feature embeddings and train other Machine Learning algorithms. Finally those algorithms were validated and tested with the segments that were previously identified by the Bird Song Detector as containing bird songs. After applying this pipeline, we were able to greatly improve the classification of bird species from Doñana present in each segment. Classifications using the detector first vs. those only using BirdNET, the state-of-the-art approach, improved: True Positives were increased, and False Negatives were reduced.}
    \label{fig:proposed_pipeline}
\end{figure*}

In the case of bird vocalization recognition, Deep Learning techniques have represented a revolution for monitoring bird populations \citep{xie2023review, stowell2022computational}. One of the most popular models of bird vocalization recognition is BirdNET \citep{kahl2021birdnet}, which has proved to be successful in many cases \citep{perez2023birdnet, wood2024guidelines, schuster2024evaluation}. These models were tested in environments where the base model was especially well trained, mainly in Northern America and central-Northern Europe. This means that the model was initially trained on a dataset that closely resembles the conditions of the test environment, thereby increasing its predictive accuracy. While BirdNET performs well in regions it was trained on, its accuracy declines in unfamiliar soundscapes due to local variations in bird vocalizations, background noise, and overlapping bird vocalizations \citep{Beery_Efficient_Pipeline_for, lauha2022domain, perez2023birdnet}. In these cases, False Positives (FPs) often arise from other vocalizing animals not being birds, anthropogenic sounds, or weather conditions \citep{stowell2019automatic, kahl2021birdnet, clark2023effect}.

Foundational models like BirdNET also struggle to recognize species they were not trained on. In theory, it is possible to retrain an existing model to add missing species, known as fine-tuning \citep{improvingmachinelearning_lalor}. Alternatives such as Google's Perch model \citep{hamer2023birb} and custom models based on BirdNET with fine-tuning for local conditions have been developed \citep{brunk2023quail, sossover2024using, ghani2024generalization, perez2025optimisation}. These fine-tuned models aim to improve accuracy by accounting for the unique characteristics of local bird populations \citep{lauha2022domain}. However, this task is quite challenging as it requires Machine Learning expertise similar to having to train models from scratch. While recent studies have demonstrated progress in applying Machine Learning to accelerate the annotation process in ecoacoustics \citep{martin2022rookognise, sethi2024large}, these approaches are still under development and not yet widely implemented.

To address these limitations, we have been inspired by methodologies used in camera trap projects \citep{Beery_Efficient_Pipeline_for, rigoudy2023deepfaune}, in which a detector is applied prior to species classification on captured images. Similarly, we propose a two-stage pipeline approach that uses first a generalizable bird vocalization detector combined later with a classifier, improving accuracy and reducing false positives. In particular, we have developed a Bird Song Detector based on the YOLOv8 model (You Only Look Once v8; \cite{Jocher_Ultralytics_YOLO_2023}) trained with data from our case study in Doñana National Park. Once the audio segments containing bird songs have been detected, we applied several classifier models, and in all of them, we have proven an improvement of the classification when the detector was present (Figure \ref{fig:proposed_pipeline}).

The added value of this pipeline lies in its ability to isolate relevant segments of audio that contain bird vocalizations before applying a more computationally intensive species classifier. This process not only reduces the number of non-bird segments incorrectly identified as bird vocalizations but also simplifies the fine-tuning of the species classifier, as the model is optimized to work with segments that are already confirmed to contain bird sounds. By separating the detection and classification stages, we minimize background noise interference and focus the classifier's resources on the most relevant audio data, leading to improved overall system performance. The generalization ability of the Bird Song Detector also ensures that it can identify bird vocalizations for locally abundant species for which it was not originally trained, making this approach robust for real-world applications with diverse species compositions. 

\begin{figure*}[ht]
    \centering
    \includegraphics[width=1\linewidth]{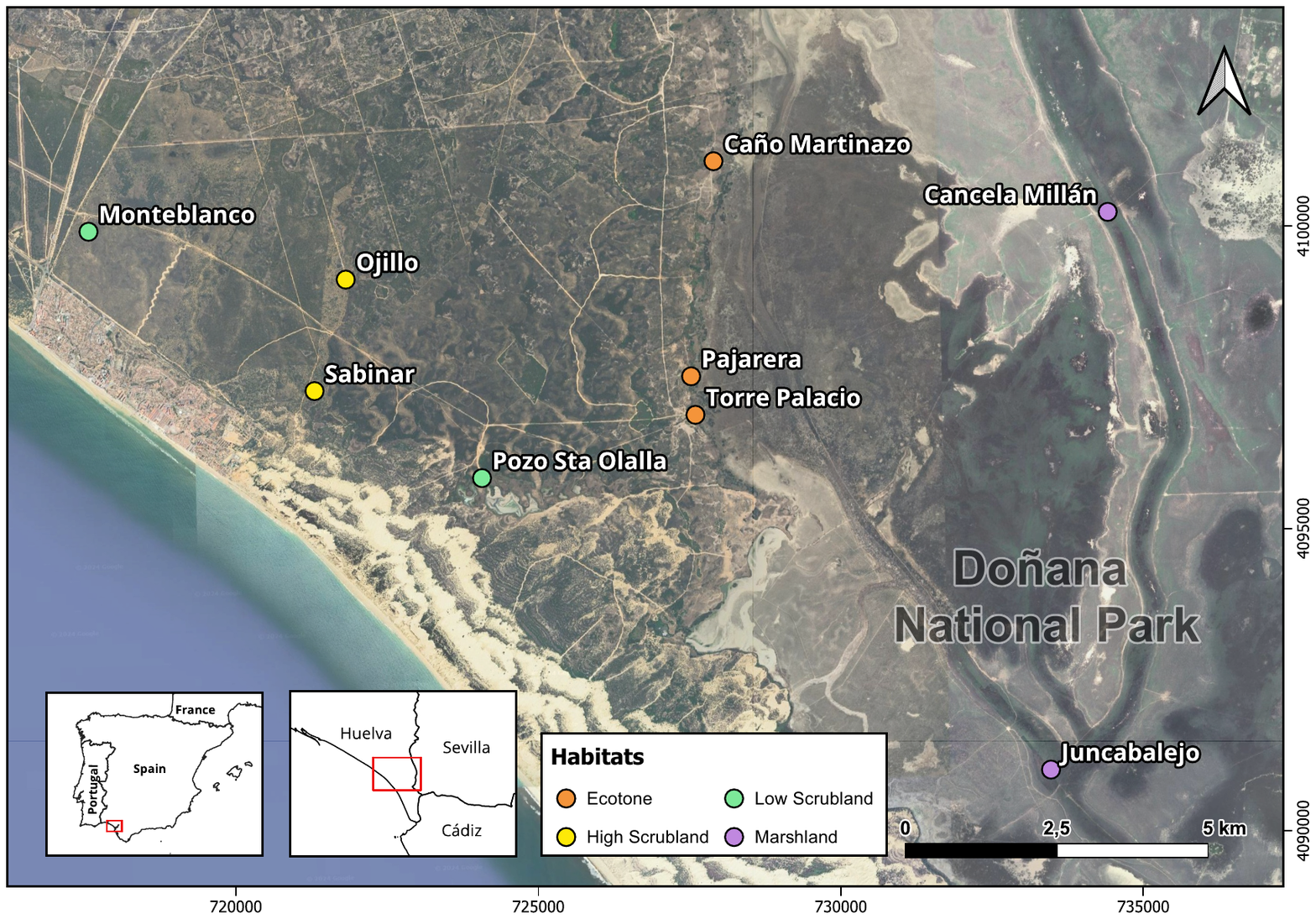}
    \caption{Spatial distribution of the nine sampling sites where PAM devices (AudioMoths) were deployed in Doñana National Park. The sampling design included the three main habitats of Doñana: marshland, scrubland (high or low), and ecotone. Each habitat type is distinguished by a different colour in the map. Names of study sites belong to the local denomination by the ICTS-Doñana \citep{ictsdonana} permanent infrastructure where the devices were installed.}
    \label{fig:sampling_sites}
\end{figure*}
As a study case, we applied this pipeline to the PAM program developed by the BIRDeep project at Doñana National Park (SW Spain) \citep{birdeeporg}. This area presents a high conservation concern due to the alarming decline of its bird populations over time due to different human impacts. Monitoring bird diversity through PAM serves then as a proxy for assessing the health status of Doñana's endangered ecosystems. 

\section{Material and Methods}
\label{material_and_methods}

\begin{comment}
In this section, we provide a detailed description of the materials and methods employed in our study. We begin by discussing the field site and acoustic data collection process, followed by the data preparation steps necessary for transforming the raw audio into a format suitable for analysis. Next, we delve into the model development, highlighting the methodologies and techniques used to build and fine-tune our models. Finally, we describe the data distribution strategy and outline our approach for model evaluation to ensure robust performance assessment.
\end{comment}

\subsection{Preprocess}
\subsubsection{Field study site and PAM design}

Soundscapes were recorded at Doñana National Park (SW Spain). This area corresponds to the marshes of the Guadalquivir delta and is one of the most important wetlands in Southern Europe, where millions of migrating birds stopover and winter every year \citep{rendon2008status, green2016donana}. The area has serious conservation threats due to water over-exploitation for agriculture and tourism, climate and land-use change, and pollution due to a mine spillover in 1998, among others \citep{green2024groundwater}, which poses serious concerns both at the local and European administration levels. These threats are particularly affecting bird populations, which are in decline over the last years \citep{camacho2022groundwater, campo2022assessing}. Doñana includes four major habitats, which are differentiated by their flooding regime and vegetation: coastal dunes, scrublands, marshlands, and the ecotone or transition among them. The deployment design of the BIRDeep project included nine AudioMoth recorders \citep{hill2019audiomoth} that were distributed among three of these habitats: two in the marshland, three in the ecotone and four in the scrubland, differentiating high and low scrubland (see Figure \ref{fig:sampling_sites}). AudioMoths are low-cost automatic audio recording devices with open-source hardware \citep{hill2018audiomoth}. They continuously recorded 1 minute of audio every 10 minutes. Configuration parameters of deployed AudioMoth included a sampling rate of 32 kHz, a medium gain, and a filter band focused on bird frequencies (0.6-16.0 kHz).

\subsubsection{Acoustic data annotation by experts}

Although devices have been continuously recording since their deployment, for this study, we selected a manually annotated subset spanning from March to October 2023 for logistic reasons. From the total of audios, we selected 461 minutes. We tried to balance representation across sites and habitats while keeping the annotation effort manageable. We also prioritized  periods of high bird activity, primarily the morning chorus, to maximize the number of detectable vocalizations in the annotations \citep{robbins1981effect}.

Annotation was carried out manually by two co-authors with ornithological expertise (ESG and GB). The process is labor-intensive and time-consuming, requiring simultaneous listening to the audio and visualization of the spectrogram, which displays how the sound's frequency content evolves over time. Annotating these 461 minutes required approximately 13,445 minutes (about 224 hours) of labor effort, with a median annotation time of 18 minutes and an average of 29.4 minutes per 1-minute file.  The experts used Audacity software \citep{audacity2017audacity}, which facilitates species identification using auditory signals and visual patterns based on spectrograms. These annotations were then exported from the Audacity format and converted to CSV files. Each annotation consisted of bounding boxes with the minimum and maximum frequency, as well as the start and end time of each bird vocalization in the spectrogram.

When ornithological experts faced uncertainties for labeling species in the audios, they referred to field censuses done in the same sampling stations to ensure the accuracy of their annotations. These field censuses were conducted periodically (43 censuses from March 2023 until February 2024) and provided diversity data that was used to cross-validate the audio labels. By cross-validating the audio data with field observations, the annotators could confirm species presence and improve the reliability of their annotations at the same time that it helped to narrow down the number of potential species, making it easier to identify the species present in the recordings.

\begin{figure*}[!ht]
    \centering
    \begin{subfigure}[b]{0.49\textwidth}
        \includegraphics[width=\textwidth]{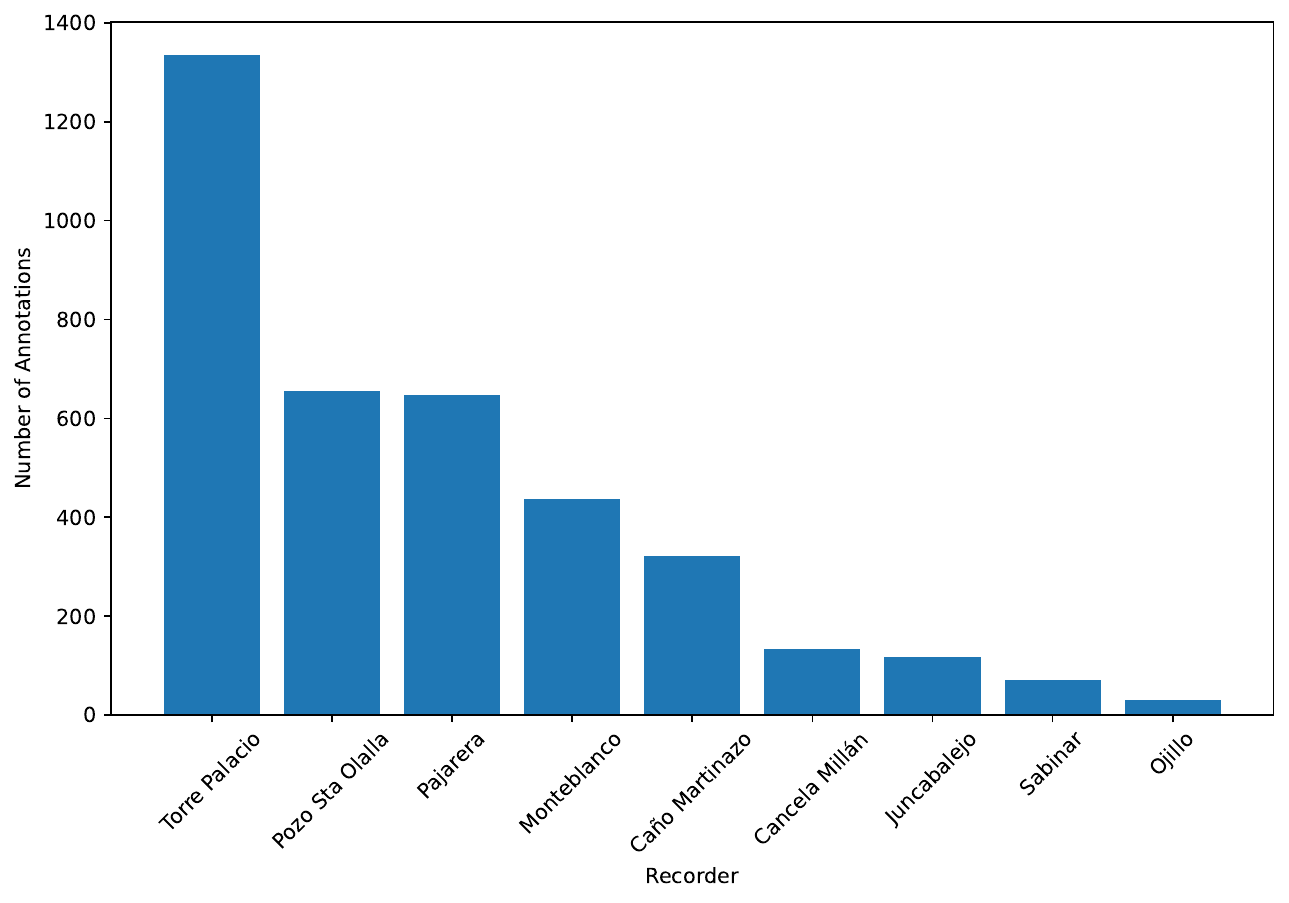}
        \caption{Total number of vocalization annotations per recorder.}
        \label{fig:audios_by_recorder}
    \end{subfigure}
    \hfill
    \begin{subfigure}[b]{0.49\textwidth}
        \includegraphics[width=\textwidth]{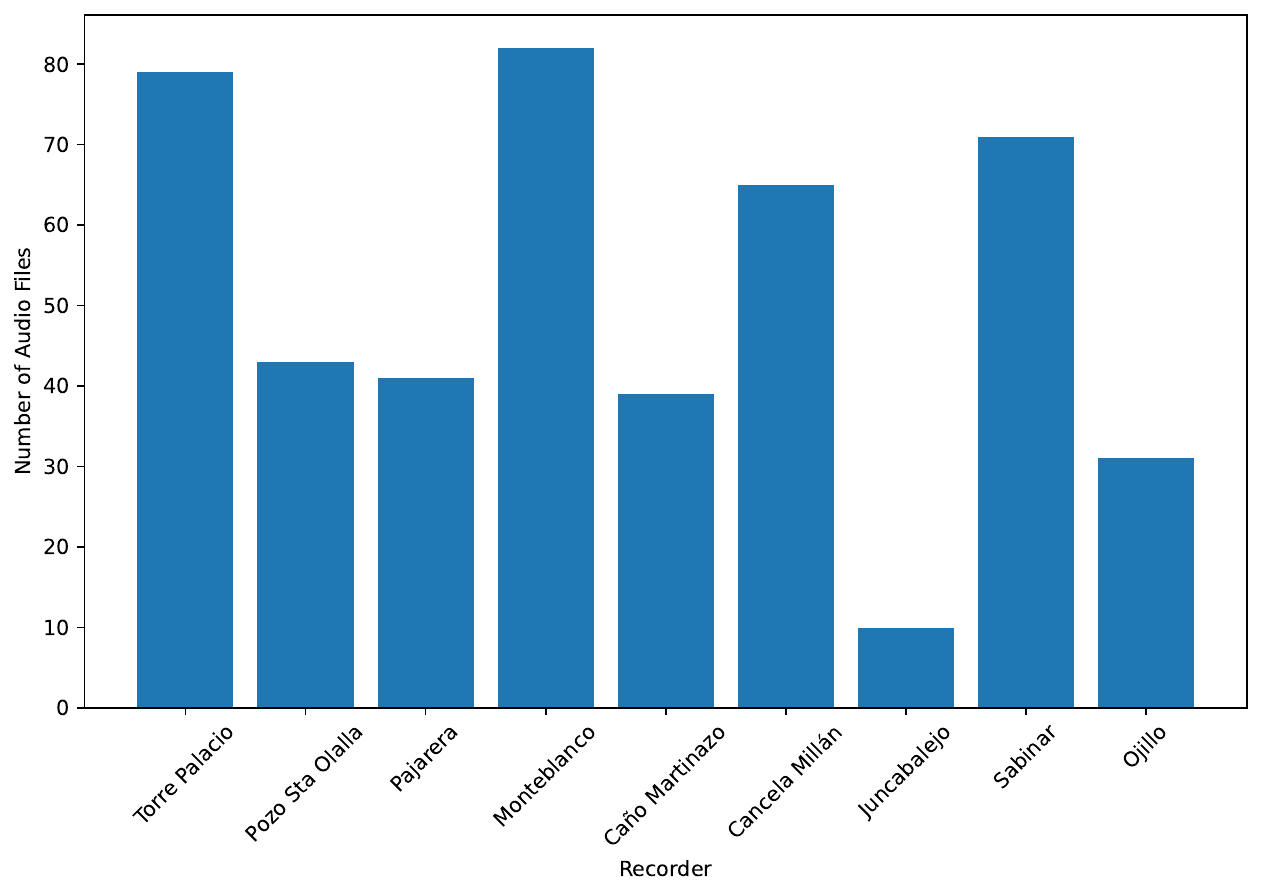}
        \caption{Number of unique audio files per recorder.}
        \label{fig:annotations_by_recorder}
    \end{subfigure}
    \caption{Number of annotations and audio files across sampling sites (see Figure \ref{fig:sampling_sites} for reference of site names), illustrating the varying levels of annotation effort across recorders.}
    \label{fig:frequency_by_recorder}
\end{figure*}

The number of annotations was highly unbalanced across recorders, whereas the number of annotated files was more similar, except for \textit{Juncabalejo} site (Figure \ref{fig:frequency_by_recorder}). This was the most isolated site in the marshland and we found logistic problems to access the recorders to change batteries and SIM cards during the flooding time. Some recorders had a high number of annotations because their recordings were rich in bird vocalizations, likely reflecting variations in bird occurrence and activity, environmental conditions, and recording quality across sampling sites.

\begin{figure*}[!ht]
    \centering
    \includegraphics[width=1\linewidth]{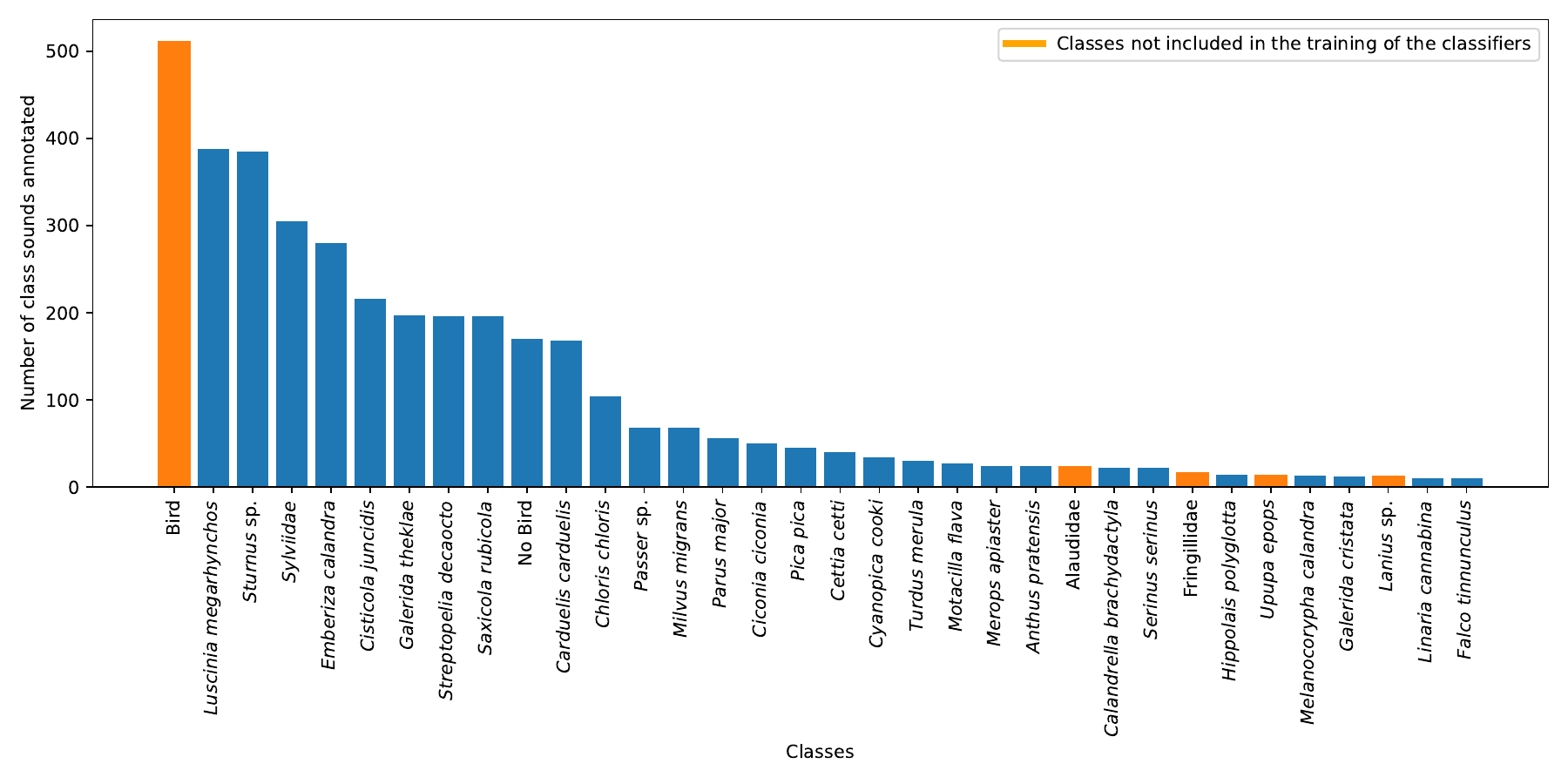}
    \caption{Number of annotations of the 34 annotated classes in the dataset. All annotated classes, including genus- and family-level labels (e.g., \textit{Sturnus} sp., \textit{Fringillidae}) and the general \textit{Bird} label, were used to train the Bird Song Detector by relabeling them under a unified \textit{Bird} class. For species classification, only the 29 classes shown in blue were retained. The orange bars represent general or ambiguous classes (e.g., family-level or uncertain identifications) that were excluded from classifier training to ensure specificity.}
    \label{fig:species_distribution}
\end{figure*}

After standardizing the annotations, a total of 3749 annotations were created, spanning 34 different classes, as shown in Figure \ref{fig:species_distribution}. In addition to the species-specific classes, we have distinguished other general classes: \textit{Sturnus} sp., \textit{Passer} sp., and \textit{Lanius} sp., which were used when the species was unknown but the genus could be identified; \textit{Alaudidae}, \textit{Fringillidae}, and \textit{Sylviidae}, used when only the bird family could be determined; a general \textit{Bird} class for cases where the sound was identified as avian but further classification was not possible. Finally, there was a \textit{No Bird} class for recordings that contain soundscapes without bird songs or any non-avian biotic sound. For the Bird Song Detector, which only distinguises between two classes \textit{Bird} and \textit{No bird}, all bird-related classifications were re-labeled as \textit{Bird}. For the species classifiers, to avoid confusion, general classes that overlapped with specific species (e.g., \textit{Alaudidae}, \textit{Bird} and \textit{Fringillidae}) were removed. \textit{Upupa epops} was also removed, due to split considerations, explained in next Section \ref{subsubsection:dataset_split}. These classes, represented in orange in Figure \ref{fig:species_distribution}, were excluded to ensure a clear distinction between general and specific labels.

It is important to note that the dataset \citep{alba_marquez_rodriguez_2024} exhibits class imbalance, with varying frequencies of annotations across different classes. Additionally, the dataset contains inherent challenges related to environmental noise (see subsection \ref{training}).

\subsubsection{Dataset split}
\label{subsubsection:dataset_split}

\begin{figure*}[ht!]
    \centering
    \includegraphics[width=1.018\linewidth]{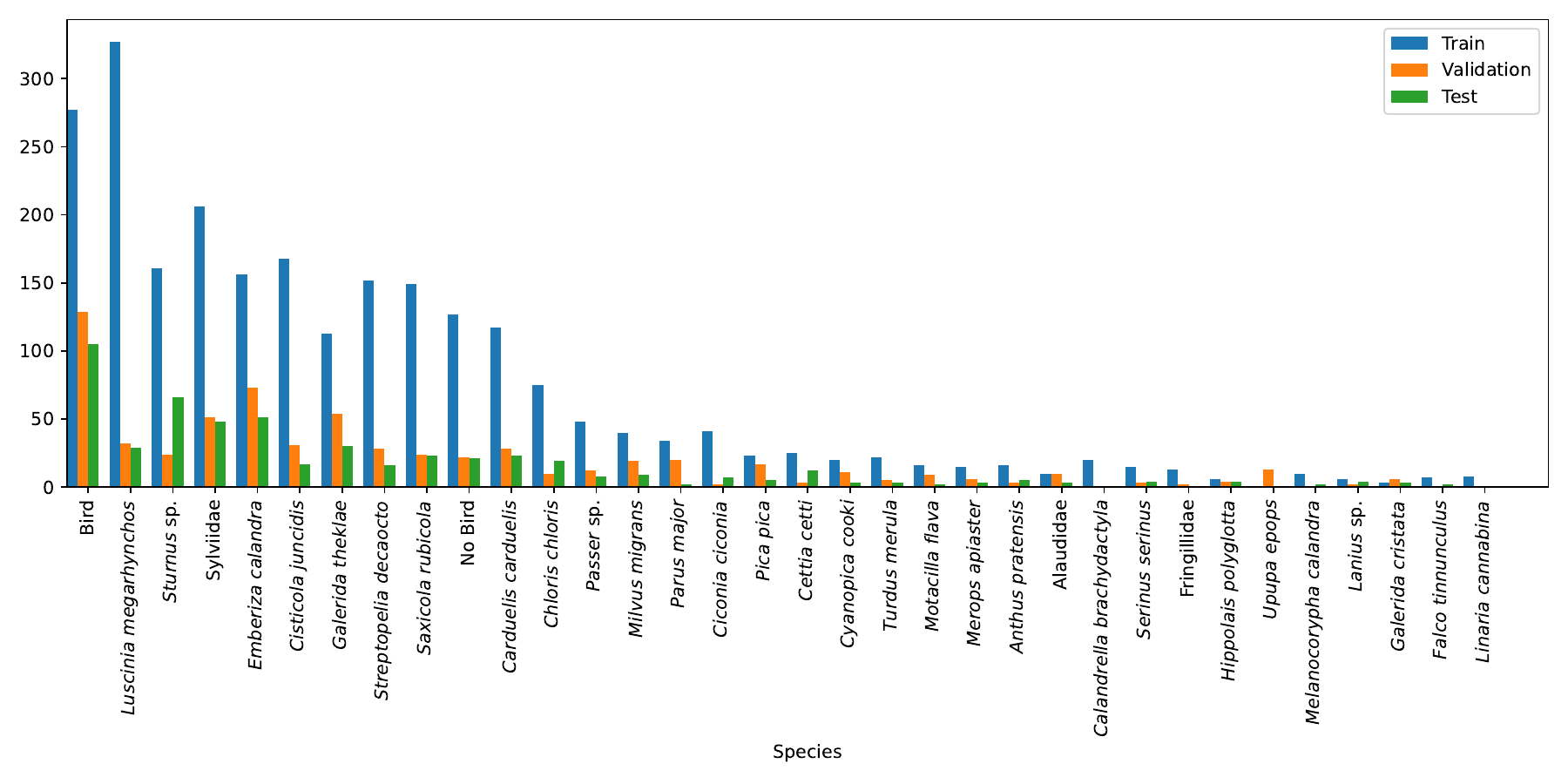}
    \caption{Distribution of the number of annotations per class across training (blue), validation (orange), and test (green) sets.}
    \label{fig:data_distribution}
\end{figure*}

The dataset \citep{alba_marquez_rodriguez_2024} was divided into training, validation, and test sets, with the aim of achieving an 80-10-10 proportion per species \citep{hardy2010pareto}. However, maintaining independence and avoiding correlation among subsets to avoid overestimation during model evaluation was challenging \citep{kattenborn2022spatially}, as some audios were multilabeled and contained vocalizations from more than one species (see Figure \ref{fig:frequency_by_recorder}). This made it difficult to strictly adhere to the desired 80-10-10 ratio. To mitigate these issues, we prioritized ensuring that no audio file appeared in more than one subset, even if it contained multiple species, to maintain independence. In the particular case of \textit{Upupa epops}, all vocalizations of this species were contained within a single audio file  assigned to the validation set, which also included annotations of other species. Consequently, \textit{Upupa epops} was placed exclusively in the validation set and excluded from both training and testing tests to prevent bias during detector evaluation. This species was not included in the classifier due to the lack of additional audio recordings (Figure \ref{fig:species_distribution}). The final distribution of sets reflects these adjustments that balance classes as much as possible but also consider independence constraints (Figure \ref{fig:data_distribution}).

\subsubsection{Data Preparation}

The audio data were transformed into spectrograms, which are graphical representations that display how the signal's energy is distributed across different frequencies over time. This transformation makes the application of Deep Learning models based on image processing techniques, i.e. Convolutional Neural Networks, suitable for audios \citep{carvalho2023automatic}. A log-scaled spectrogram is a variant of a spectrogram where the frequency axis is represented logarithmically, emphasizing lower frequencies while preserving the visibility of higher frequencies. This scaling is particularly suitable for analyzing bird vocalizations, as it provides greater resolution in the lower frequency range where many bird songs and calls are concentrated, while still capturing harmonics and other features that extend into higher frequencies \citep{wyse2017audio}. 

\begin{figure*}[!ht]
    \centering
    \includegraphics[width=1.01\linewidth]{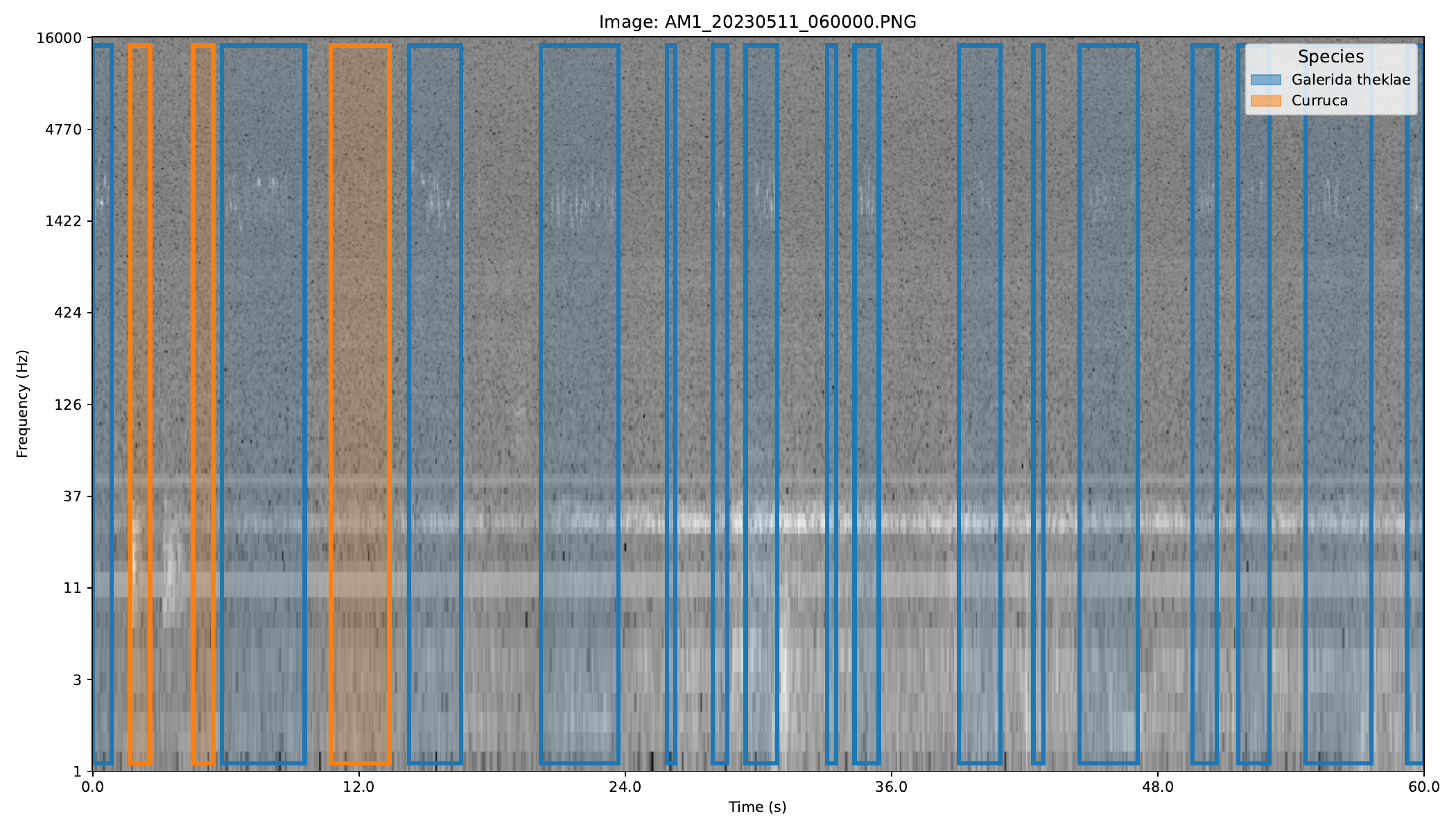}
    \caption{An example of a log-scaled spectrogram, in gray scale for clearer visualization, from the Doñana dataset with annotations (i.e., blue and orange rectangles) for temporal windows and the complete frequency spectrum for the annotated vocalizations of two bird species.}
    \label{fig:annotatedspectrogram}
\end{figure*}

Spectrograms were generated using the \texttt{librosa} Python library \citep{librosa2025}, a popular toolkit for audio analysis. The Short-Time Fourier Transform (STFT) was applied to the audio signal to compute the spectrogram, and the amplitude of the resulting frequency bins was converted to a decibel (dB) scale. The frequency axis was displayed on a logarithmic scale. The frequency range of the spectrograms was set to a minimum of 1 Hz and a maximum of 16,000 Hz, which encompasses the typical range of bird vocalizations while excluding inaudible frequencies or irrelevant noise. The dimensions of the output spectrogram images were $930\times462$ pixels. Although our annotations originally included frequency windows for each vocalization, to simplify detection given the limited data, bounding boxes spanned the entire frequency y-axis. This choice prioritizes the temporal localization of bird vocalizations, which are the primary signal for subsequent classification (see Figure \ref{fig:annotatedspectrogram} as an example).

To enhance the dataset and improve the robustness of downstream models, additional preprocessing steps were applied. Specifically, synthetic audio samples were generated by adding random noise (using \texttt{NumPy} Python library \citep{harris2020array}, to simulate background interference and slightly adjust the intensity of the signals. These modifications aimed to increase the number of training examples that contained only background noise, ensuring a more diverse dataset for both detection and classification tasks.

The annotations were further processed to adopt the format required by the YOLOv8-based object detection model \citep{Jocher_Ultralytics_YOLO_2023}. A detailed explanation of the mathematical conversion from YOLOv8's normalized coordinates to temporal annotations in seconds is presented in \ref{appendix:additional_info}.

\subsection{Bird Song Detector Model Development}

We chose a pre-trained YOLOv8 model to develop our Bird Song Detector because it is the state-of-the-art for real-time object detection models.  YOLOv8's architecture consists of multiple convolutional layers followed by fully connected layers that predict bounding boxes, objectness scores, and class probabilities for detected objects. YOLOv8 divides each input image into a grid and predicts bounding boxes for the presence of these image events, along with a confidence score for each prediction. In our case study, the input images were spectrograms, and the relevant events were bird vocalizations represented as sonograms in the images. Given the nature of our local audio data, which contains a mix of bird sounds and background noise, we chose YOLOv8s (small version of YOLOv8) for distinguishing relevant events. This reduced version was more effective in maintaining high detection accuracy while minimizing computational load at initial experiments \citep{pasupa2016comparison}. 

The training of the Bird Song Detector model was done using the previously described spectrograms that were annotated with the bird vocalizations from Doñana. We performed data augmentation techniques to improve the robustness of the model: YOLO’s internal augmentation methods, which involve modifications such as changes in HSV (hue, saturation, and value), temporal translations (shifting the audio in time), and \textit{mixup} \citep{zhang2017mixup, tokozume2017learning}, a technique widely used in audio data augmentation. The dataset was also supplemented with additional samples from an external dataset \citep{piczak2015esc} to further enrich model training (see Section \ref{training} below).

\subsubsection{Evaluation metrics for Detections}

Currently, there are no widely available, general-purpose bird vocalization detectors trained for global-scale datasets. As a result, a direct comparison of our Bird Song Detector with another state-of-the-art bird vocalization detector is not possible. Existing tools like BirdNET, while primarily designed for species classification, also includes background training data. Therefore, BirdNET can serve as a baseline detector for identifying audio segments containing bird vocalizations and we compared it with the Bird Song Detector developed in our study. More information about BirdNET is provided in Section \ref{section:birdnet_classifier}.

The performance of both detectors used (Bird Song Detector and BirdNET) was assessed using standard metrics for object detection and classification tasks, including TPs, FPs, FNs, Precision, Recall, F1-score, and Accuracy. The detection metrics were calculated as follows:
\begin{equation*}
\text{Precision} = \frac{\text{TP}}{\text{TP} + \text{FP}}
\end{equation*}

\begin{equation*}
\text{Recall} = \frac{\text{TP}}{\text{TP} + \text{FN}}
\end{equation*}

\begin{equation*}
\text{F1-Score} = \frac{2 \cdot \text{Precision} \cdot \text{Recall}}{\text{Precision} + \text{Recall}}
\end{equation*}

\begin{equation*}
\text{Accuracy} = \frac{\text{TP} + \text{TN}}{\text{TP} + \text{FP} + \text{FN} + \text{TN}}
\end{equation*}

Additionally, we applied an Intersection over Union (IoU) criterion to ensure that detected segments overlapped with Ground Truth annotations, making the evaluation more consistent across different detection approaches. A prediction is considered correct if its overlap with a Ground Truth  annotation meets the required threshold, quantified using the IoU metric. IoU measures the intersection between predicted and actual vocalization segments relative to their total duration. It is defined as:

\begin{equation*}
\text{IoU} = \frac{\text{Intersection (Overlap Duration)}}{\text{Union (Total Duration of Prediction and Ground Truth)}}
\end{equation*}

\paragraph{Bird Song Detector}

For this study, a minimum IoU threshold of 0.1 was used to determine matches. Predictions with IoU values below this threshold were classified as FP, and Ground Truth annotations without matching predictions were classified as FN. Temporal segments from the Bird Song Detector were further filtered based on a confidence score threshold to exclude low-confidence predictions.

\paragraph{BirdNET as a bird vocalization detector}
\label{bird_vocalization_detection_comparison}

BirdNET processes audio recordings by segmenting them into fixed 3-second intervals and classifying each segment (see Section \ref{section:birdnet_classifier}). In this study, we assessed BirdNET’s ability to function as a binary bird presence detector, independent of species classification. A detection was considered correct if any part of a GT annotation overlapped with a BirdNET segment, even if the GT annotation spanned multiple segments. This approach accounts for cases where annotated bird vocalizations extend across adjacent 3-second intervals.

To ensure a fair comparison with the Bird Song Detector, we applied an IoU-based evaluation. Because BirdNET operates on fixed 3-second windows, its segments do not always align precisely with ground truth annotations. Therefore, instead of a strict IoU match, we required a minimum overlap between GT annotations and predicted segments, similar to the IoU threshold used for the Bird Song Detector.

BirdNET assigns a confidence score to each prediction, reflecting the likelihood that a species vocalization is present. To ensure a species-agnostic evaluation, we applied a uniform confidence threshold across all predictions rather than optimizing it for individual taxa. The selected confidence threshold determines how conservatively BirdNET predicts the presence of a bird vocalization:
\begin{itemize}
    \item \textbf{Confidence Score = 0.1 (Default Value)}: This is the default BirdNET threshold. Prior studies have shown that using this threshold in real-world PAM scenarios leads to an increase in FPs, as BirdNET assigns low-confidence scores to background noise or ambiguous detections \citep{perez2023birdnet}. However, lowering the confidence threshold improves recall by reducing missed detections.
    \item \textbf{Confidence Score = 0.6 (Adjusted Threshold)}: Higher confidence thresholds have been recommended for improving the reliability of BirdNET predictions in complex acoustic environments. \cite{funosas2024assessing} found that BirdNET predictions become significantly more reliable when using thresholds above 0.7. We adopted a slightly lower threshold of 0.6 to balance precision and recall, reducing FPs while still capturing a sufficient number of detections.
\end{itemize}

To systematically evaluate BirdNET as a detector, we applied both 0.1 and 0.6 thresholds uniformly to all species, treating BirdNET’s predictions as binary vocalization detections rather than species-specific classifications. This ensures that confidence score variations across species do not influence the detection performance comparison.

Different evaluations were carried out with different species lists. First, we used a full list of species from Doñana that included 412 classes, as provided by BirdNET, based on the study area coordinates. We then later reduced this list by selecting the most common species in the area and removing sporadic observations, using references from existing literature \citep{garcia2000prontuario, garrido2004anuario} and our own field observations from expert censuses. This resulted in a shorter list of 337 species, which we have called the ``expert list''.

To comprehensively assess BirdNET’s performance under different conditions, we tested the following four configurations:

\begin{itemize}
    \item \textbf{BirdNET without fine-tuning at a confidence threshold of 0.6}, using the default species list for Doñana National Park coordinates.
    \item \textbf{BirdNET without fine-tuning at a confidence threshold of 0.6}, using the expert species list for Doñana National Park.
    \item \textbf{BirdNET fine-tuned at a confidence threshold of 0.6}, using classifier species classes.
    \item \textbf{BirdNET fine-tuned with a lower confidence threshold of 0.1}, to assess detection sensitivity.
\end{itemize}

\paragraph{Comparison of Methods}

The percentage improvements in TPs, FNs, FPs were calculated based on changes observed between both methods, using the following general formula:

\begin{equation*} 
\text{\textit{PercentageChange}} = \left( \frac{\text{\textit{NewValue}} - \text{\textit{OldValue}}}{\text{\textit{OldValue}}} \right) \times 100 
\end{equation*}

Where $ \textit{NewValue} $ refers to the value obtained from our Bird Song Detector and $ \textit{OldValue} $ refers to the value obtained from BirdNET with the specific confidence score threshold.

An increase in TPs indicates an improvement in detection accuracy, as more bird vocalizations are correctly identified. A decrease in FNs is also a sign of improved performance, as the system misses fewer bird vocalizations. Conversely, a decrease in FPs represents a reduction in erroneous detections of non-bird sounds. For all metrics, positive percentage changes in TPs and negative changes in FN and FP signify improvements, while negative changes in TPs or positive changes in FNs and FPs indicate a decline in performance.

When evaluating percentage changes in the results, it is important to consider them in absolute terms. A large percentage increase or decrease might appear significant at first glance, but its actual impact depends on the scale of the values involved. For instance, changes from small baseline values can result in high percentage variations, even if the absolute difference is relatively minor. Conversely, changes in metrics with larger absolute values might show smaller percentage shifts but represent a more substantial impact on overall performance. Therefore, it is crucial to interpret these percentages within the context of the absolute figures to avoid misinterpreting the true extent of the changes.

\subsection{Bird-Song Classifier Model Development}

\subsubsection{BirdNET}
\label{section:birdnet_classifier}

As the third step in our pipeline (Figure \ref{fig:proposed_pipeline}), we fine-tuned BirdNET v2.4 to create a feature extractor specifically adapted to the ecological context of Doñana. BirdNET is a Deep Learning model designed to classify bird species using audio inputs. It segments audio recordings into 3-second clips, it transforms the audio into spectrogram images, and it performs the classification into a bird species using a deep Convolutional Neural Network \citep{kahl2021birdnet}. BirdNET v2.4 uses an EfficientNetB0-like backbone with a final embedding size of 1024 for feature extraction and classification. It covers frequencies from 0 Hz to 15 kHz, and it is trained for 6,522 classes (including 10 non-event classes), making it suitable for the identification of diverse bird species all over the world. Non-event classes refer to categories that represent sounds or signals that are not related to bird vocalizations, such as background noise, human-made sounds, or other environmental noises.

The fine-tuning was performed in the BirdNET v2.4 GUI v1.5.1 \citep{BirdNET-Analyzer}, with the default training parameters. The training mode was set to \texttt{replace}, ensuring that the original classification layer was overwritten and only the newly trained classes remained. To address class imbalance, \texttt{repeat} upsampling was applied, with minority classes resampled at 10\% of the majority class frequency. This fine-tuning step adapted BirdNET to the bird species and soundscapes of Doñana, intending to reduce bias and improve model performance for audio recordings from the region.

Once fine-tuned, the model was used to extract feature embeddings—1024-dimensional vector representations capturing essential audio characteristics. These embeddings served as input features for subsequent classifiers.

The training data for fine-tuning were derived directly from the expert annotations \citep{alba_marquez_rodriguez_2024}. Each annotated time window was segmented and organized into folders corresponding to their respective annotated classes, adhering to BirdNET's required training input structure.

\subsubsection{Other classifiers}
\label{subsub:other_classifiers}

In addition to the fine-tuned BirdNET model, we explored other classification approaches: a Machine Learning classifier, and two Deep Learning models.  

For the Machine Learning approach, we used Random Forest, a well-established method in bioacoustics for species classification due to its ability to handle structured data efficiently and provide stable predictions \citep{breiman2001random}. The model was trained using BirdNET embeddings (1024-dimensional feature vectors) extracted from the same dataset used for fine-tuning. 

The Random Forest model was implemented using the \texttt{RandomForestClassifier} from the \texttt{scikit-learn} library \citep{scikitlearn} of Python. The optimal configuration included a maximum depth of 50, with a minimum of 2 samples per leaf and 10 samples per split.

For Deep Learning models, we trained ResNet50 and MobileNetV2 using the \texttt{Keras} Python library \citep{chollet2015keras}, initializing both with pre-trained \textit{ImageNet} weights. The original classification layer was removed, and all layers were initially frozen. A new classification head was added, consisting of a Global Average Pooling layer, followed by a 256-unit dense layer with ReLU activation and a final softmax layer for classification. The model was first trained with only the newly added layers, keeping the pre-trained base frozen. After this initial phase, the entire model was unfrozen for a final fine-tuning step across all layers. MobileNetV2 was selected due to its reduced dimensionality, which helps to prevent overfitting, especially given the relatively small dataset size, while also reducing training time and computational requirements. Both deep learning models were trained using 3-second spectrograms generated from the segments used for the fine-tuned BirdNET.

During evaluation, the models were tested using both full 1-minute audio files (segmented into 3-second windows) and shorter audio segments identified by the Bird Song Detector. When these detected segments were shorter than the required 3-second input length, they were padded with zeros—representing silence—to match the model’s input size.

We compared these four classifiers (fine-tuned BirdNET, Random Forest, and two Deep Learning models) with and without previously using the Bird Song Detector. This allowed us to evaluate how the Bird Song Detector improved bird classification in different approaches.

\subsubsection{Evaluation metrics for Classifiers}

All BirdNET-based classifiers were evaluated using a fixed confidence score threshold of 0.1, both with and without the Bird Song Detector. If BirdNET did not assign any species a confidence score above this threshold, the segment was classified as background. However, for other classifiers trained specifically for this study (e.g., Random Forest, ResNet50, MobileNetV2), a separate \textit{Background} class was included in the model itself. Since these classifiers explicitly learned to distinguish bird vocalizations from background noise, no confidence threshold was applied, and the species classification with the highest confidence score was selected for each segment, whether it was a bird species or background noise. This approach ensured that BirdNET adhered to a stricter filtering process, while other models relied on their learned classification structure.

The classification performance of the models was evaluated using standard metrics from the \texttt{scikit-learn} Python library \citep{scikitlearn}, including the confusion matrix, accuracy, precision, recall, and F1-score, along with custom indices designed to assess the classifier’s ability to estimate the total number of bird vocalizations.

Beyond standard classification metrics, we introduce the Idx Pred/Ann metric to further analyze model behavior. This metric helps evaluate not only classification accuracy but also the ability of models to filter additional false positives, providing insight into whether the classifier tends to over-predict or under-predict species occurrences:

\begin{itemize}
    \item \textbf{Number of Predictions}: The total number of predictions made by the classifier, regardless of their correctness.
    \item \textbf{Idx Pred/Ann}: Index of Predictions per Annotation. This index quantifies the degree to which the classifier overestimates (\( \text{Idx Pred/Ann} > 1 \)) or underestimates (\( \text{Idx Pred/Ann} < 1 \)) the number of bird vocalizations in the test set. A value close to 1 indicates that the classifier predicts a number of vocalizations similar to the number of annotated ground truth segments, while deviations highlight either an excessive or insufficient number of predictions. It evaluates the capability of detection and filtering additional false positives.
\end{itemize}

\section{Experiments}
\label{experiments}
  
% A Theory section should extend, not repeat, the background to the article already dealt with in the Introduction and lay the foundation for further work. In contrast, a Calculation section represents a practical development from a theoretical basis

\subsection{Bird Song Detector Training}
\label{training}

To select the Bird Song Detector, multiple YOLOv8 models with different configurations were trained, and their performance was evaluated using the mean Average Precision (mAP) metric at an Intersection over Union (IoU) threshold of 50\% (mAP50) \citep{padilla2021comparative}. Initial experiments, represented by the purple lines in Supplementary Figure \ref{fig:training} (\textit{Base}, \textit{Hyperparameter Exploration V1}, \textit{Hyperparameter Exploration V2}, \textit{AugmentedBG V1} and \textit{AugmentedBG V2}), showed suboptimal performance, particularly due to a high number of FPs (Table \ref{tab:experimental_results}).

Given the complexity and small size of the dataset \citep{alba_marquez_rodriguez_2024}, the bounding boxes, which were initially designed to delimit both the frequency spectrum and the time window, were simplified (represented by the light orange line, \textit{FullFrequencies}, in Figure \ref{fig:training}). This approach aimed to reduce the complexity of the task for the model, given the limited amount of data available.

To address the issue of the FPs, the ESC-50 dataset \citep{piczak2015dataset}, which is a large collection of 50 environmental sound classes, was introduced as background noise (negative samples). To prevent confusion, bird-related classes were removed only from ESC-50, but not from our primary dataset. When the dataset was fully included, the model primarily learned to recognize background sounds and failed to detect bird songs effectively (green line, \textit{AllESC50}, in Figure \ref{fig:training}).

Subsequently, the ESC-50 dataset was reduced to comprise only 25\% of the total training data. This adjustment led to significant improvements in the model's performance (dark orange line, \textit{Best Model}, in Figure \ref{fig:training}). This balanced approach allowed the model to better differentiate between bird songs and background noises, improving detection accuracy while minimizing FPs.

The various model configurations employed during the experimentation are summarized in Table \ref{tab:experimental_results}, which also presents the performance metrics for each configuration. The \textit{Best Model}, which employed synthetic background augmentation of noise and intensity changes and a reduced ESC50 dataset, achieved high mAP50 scores (0.29), along with balanced precision and recall. Other configurations, such as \textit{AllESC50}, displayed lower performance metrics. On the other hand, the \textit{Full Frequencies} model without the ESC50 dataset had the best performance during training, with an mAP50 score of 0.305. This highlights the importance of specific augmentation strategies and dataset choices in optimizing detection accuracy.

\subsection{Selection of confidence score threshold for the Bird Song Detector}
\label{selection_of_the_Confidence_Score}

To evaluate the confidence scores generated by the Bird Song Detector, we transformed confidence scores resulting from the output of the model into logit scores. This transformation allowed us to convert unitless confidence scores into a probability of detection, using a logistic regression \citep{wood2024guidelines}.

By converting the confidence scores into logit scores, we were able to assess the model’s performance more accurately across various probability thresholds \citep{padilla2021comparative, wood2024guidelines}. This transformation helped identify the point at which the model’s predictions were most reliable, ensuring that the selected threshold maximized True Positives while maintaining an acceptable level of False Positives. Through this method, we were able to ensure that the confidence scores reflected the actual likelihood of correct predictions, improving the interpretability and robustness of the model’s output.

The conversion from confidence scores to logit scores is based on the logistic function:

\begin{equation*}
\text{logit}(p) = \log \left( \frac{p}{1 - p} \right)
\end{equation*}

where \( p \) represents the confidence score of the detector's prediction. This transformation helps to interpret the confidence scores probabilistically, providing a more nuanced understanding of the detector's performance characteristics.

\begin{table}[!ht]
\scriptsize
\centering
\begin{tabular}{|c|c|c|c|}
\hline
\textbf{Probability Threshold} & \textbf{Logit Score} & \textbf{Confidence Score} & \textbf{TP Loss (\%)} \\ \hline
40\% & -2.75 & 0.06 & 0.00 \\ \hline
60\% & -1.78 & 0.14 & 22.08 \\ \hline
80\% & -0.58 & 0.36 & 74.35 \\ \hline
95\% & 1.30 & 0.79 & 99.03 \\ \hline
\end{tabular}
\caption{Comparison of different probability thresholds for detection with their respective logit scores, confidence scores, and TP losses.}
\label{tab:threshold_comparison}
\end{table}

\begin{figure}[!ht]
    \centering
    \includegraphics[width=1\linewidth]{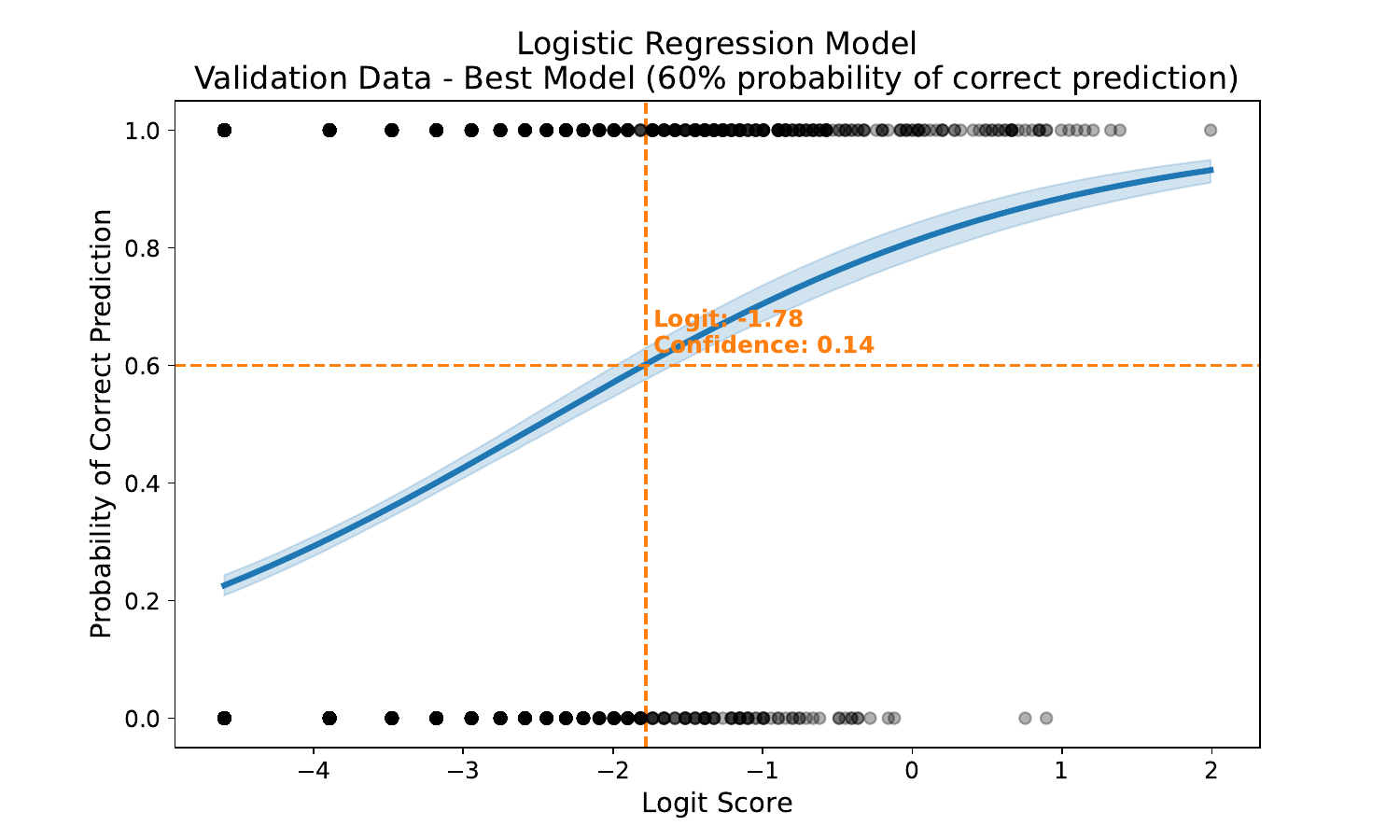}
    \caption{Logistic Regression Model with a 60\% probability threshold for accurate predictions. Each black dot (with some transparency to allow overlapping) represents a detection made by the Bird Song Detector. The confidence score of each detection is transformed into a logit score and plotted along the X-axis. A Y-axis value of 0 indicates an incorrect detection (i.e., it does not match an actual bird vocalization according to the ground truth). Conversely, a Y-axis value of 1 indicates that the detection was correct, based on the ground truth annotations provided by experts. The blue line represents the logistic regression model fitted to these data points. The shaded area surrounding the blue line represents the 90\% confidence interval (CI) for the model's predictions, calculated with a bootstrapping method. The orange lines represent the intersection of the selected threshold (60\% ) with the blue logistic regression line, showing the corresponding logit score, which is approximately -1.78 (back-transformed into a confidence score of 0.14).
   }
    \label{fig:model12_log_reg_60}
\end{figure}

We evaluated the Bird Song Detector using different probability thresholds (40\%, 60\%, 80\%, and 95\%) to find the optimal balance between maximizing True Positive (TP) and minimizing False Negatives (FN; see Table \ref{tab:threshold_comparison} and Figure \ref{fig:model12_log_reg_60}). This optimization process involved analyzing the trade-off between increasing TP detection (i.e., capturing more bird vocalizations) and keeping errors as low as possible due to FPs or FNs.  Higher probability thresholds tend to reduce the number of detections, including fewer FPs and TPs. After evaluation, we determined that the 60\% threshold provided the best balance: it minimized the loss of TPs while keeping the FN rate at an acceptable level (see Figure \ref{fig:model12_log_reg_60}). The selected threshold (60\% ) corresponds to a logit score of -1.78. When transformed back into a confidence score, this results in a threshold of 0.14. % We applied a logistic regression with a logit function to correlate the binomial status of each prediction (correct or incorrect) vs. the logit-transformed confidence score resulting from the Bird Song Detector. As expected, detections that had a higher confidence score were more likely to be correct. This line allows us to estimate the probability of a correct prediction for any given logit score, which can then be transformed back into a confidence score. We applied a bootstrapping approach, where 1000 bootstrap samples were drawn at random, and logistic regression models were fitted for each sample. This allowed us to calculate the 90\% confidence intervals of the logistic model, indicating the uncertainty in the probability estimates across the dataset (represented as a shadowed area in the Figure \ref{fig:model12_log_reg_60}). Thus, a narrower CI suggests higher certainty in the prediction, while a wider CI indicates greater uncertainty. 

In practical terms, applying a confidence score threshold of 0.14 to our Bird Song Detector ensures that for any given prediction, the probability of it being correct is at least 60\%, regardless of the original confidence score provided by the model.

Lower thresholds, such as 40\%, result in a 0\% loss of TPs but occur before the logistic regression model's effects begin to improve performance significantly (logit = -2.75, confidence = 0.06; Table \ref{tab:threshold_comparison}). At this threshold, the model fails to utilize the logistic regression adjustments effectively, as evidenced by the extremely low confidence score.

On the other hand, higher thresholds, like 80\% or 95\%, lead to excessive losses of TPs (74.35\% and 99.03\%, respectively). These thresholds result in an impractical trade-off, where the reduction in FPs comes at the cost of almost complete loss of the TPs (which need to have a high probability threshold to be retained), significantly degrading the detector’s performance.

In this case and for our case study, the 60\% threshold is high enough to ensure that the logistic regression model's adjustments are actively enhancing detection performance, while also preserving a manageable amount of TPs. This threshold ensures that the model remains both reliable and practical for detecting bird songs, thereby offering an optimal balance between confidence and detection accuracy. For further experiments and results in the Bird Song Detector, a confidence score threshold of 0.15 will be used.

% Results should be clear and concise.

\subsection{Bird Song Detector Selection}

\begin{figure}[!ht]
    \centering
    \begin{subfigure}[b]{0.45\textwidth}
        \includegraphics[width=\textwidth]{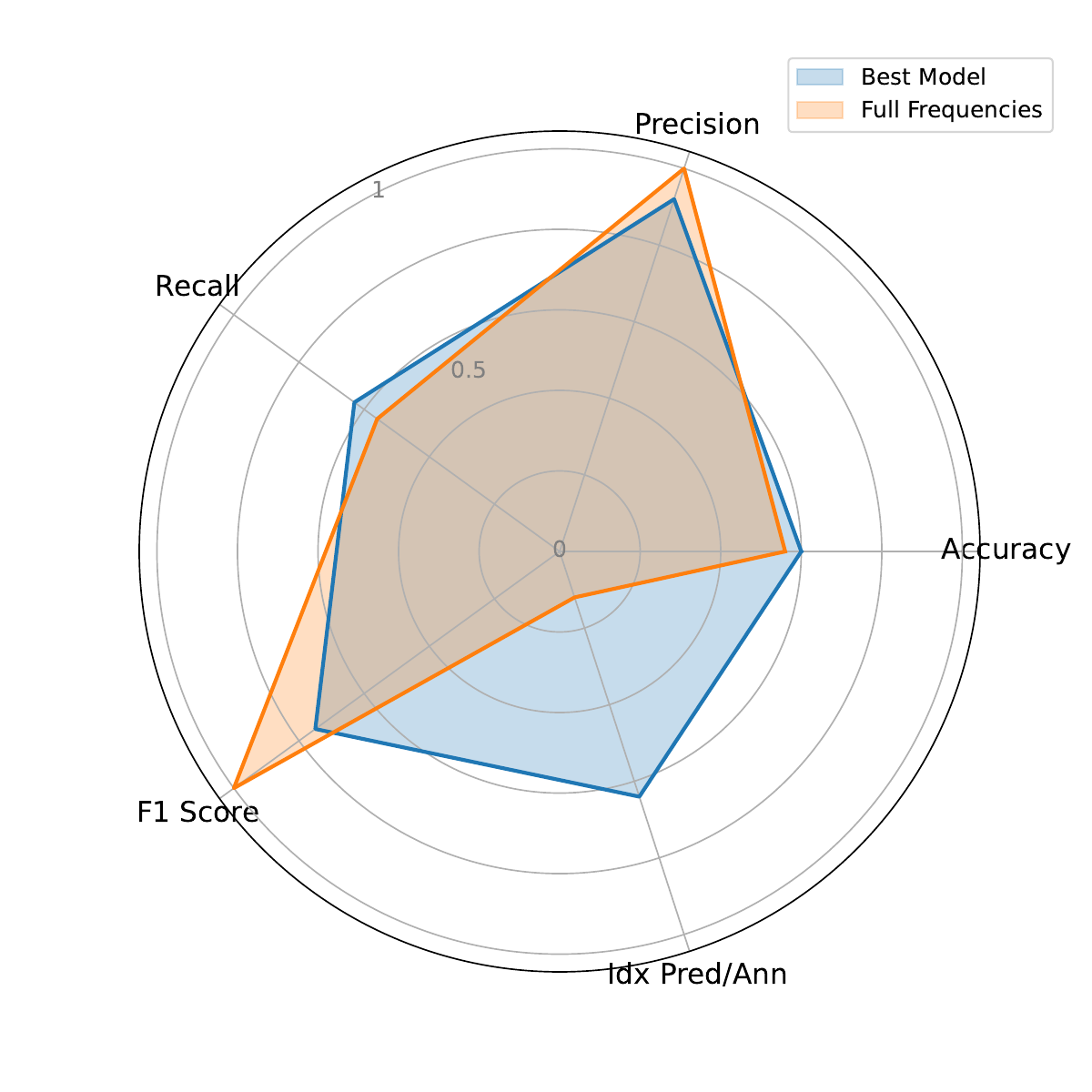}
        \caption{Comparison between models used for the Detector according to different metrics: Accuracy, Precision, Recall, F1 Score, and an index calculated based on the number of predictions relative to the total number of annotations (Idx Pred/Annn).}
        \label{fig:spider_webs_gen}
    \end{subfigure}
    \hfill
    \begin{subfigure}[b]{0.45\textwidth}
        \includegraphics[width=\textwidth]{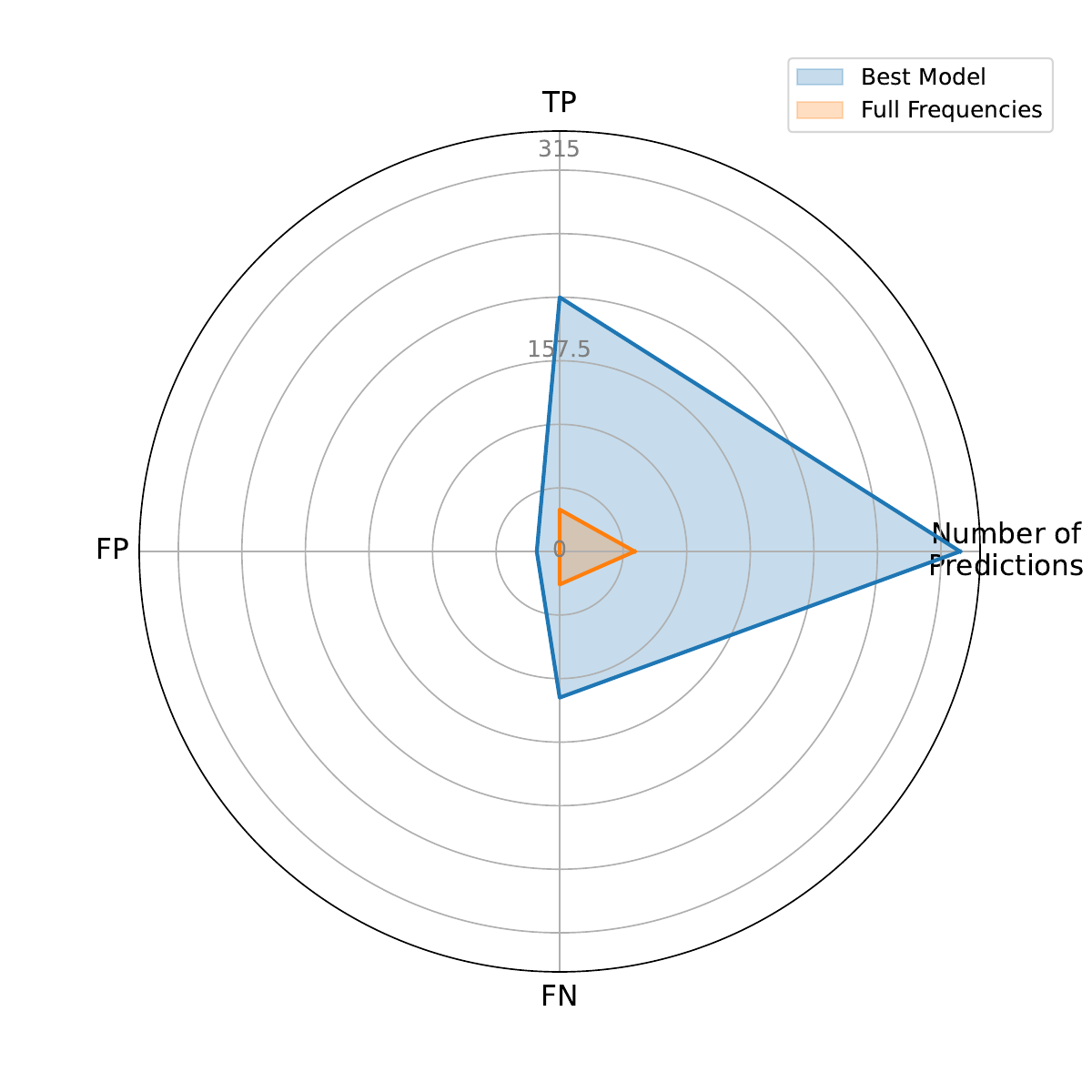}
        \caption{Detailed Breakdown of Predictions. This plot illustrates the integer values of TP, FN, FP, and the total Number of Predictions.}
        \label{fig:spider_webs_int}
    \end{subfigure}
    \caption{Spider Web Plots Comparing Model Performance Metrics. The blue area represents the \textit{Best Model}, while the orange area indicates the \textit{Full Frequencies} model.}
    \label{fig:spider_web}
\end{figure}

The performance of the top models (\textit{Best Model} and \textit{Full Frequencies}) was compared using the same validation dataset. As explained above, we set confidence scores to ensure a 60\% probability of correct predictions: 0.14 for the \textit{Best Model} and 0.06 for the \textit{Full Frequencies} model. Although both models display comparable performance across certain metrics, the \textit{Best Model} outperforms \textit{Full Frequencies}, notably in the number of predictions (315 vs. 59; Figure \ref{fig:spider_web}).  At a 60\% of correct prediction rate, the \textit{Best Model} not only generates significantly more predictions, but also maintains competitive performance metrics, including accuracy and F1-Score.

In the case of the \textit{Full Frequencies} model, although higher precision is achieved, the limited number of predictions raises concerns about its robustness and generalizability. In contrast, the \textit{Best Model}, with a greater number of predictions and balanced performance metrics, stands out as the more reliable choice for our ecological monitoring objectives, given its consistent performance across metrics. %It demonstrates more consistent performance across metrics, indicating a stable and robust detection pattern.

The \textit{Best Model} also demonstrates a better fit to the validation data and a lower prediction uncertainty, suggesting a stronger ability to generalize across the dataset and provide more reliable detections than the \textit{Full Frequencies} model. This enhanced performance can be attributed to the model's specific augmentation strategy, which includes synthetic background noise, intensity variations, and the use of a reduced ESC50 dataset. These augmentation techniques likely contribute to the model’s ability to make more confident predictions and improve overall reliability.

\section{Results}
\label{results}

\subsection{Bird Song Detector Performance}

\begin{figure}[!ht]
    \centering
    \includegraphics[width=1\linewidth]{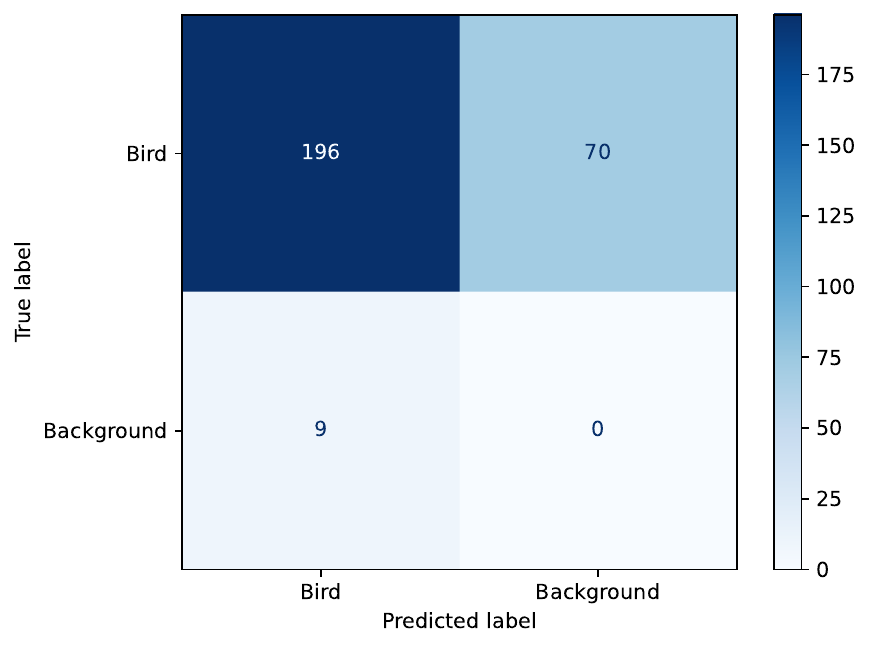}
    \caption{Binary confusion matrix for the Bird Song Detector on the test dataset with a confidence threshold of 0.15.}
    \label{fig:confusion_matrix_test_0.15}
\end{figure}

The Bird Song Detector built with the \textit{Best Model} was evaluated on the test dataset, and its binary confusion matrix for these detections is shown in Figure \ref{fig:confusion_matrix_test_0.15}. This matrix shows that the detector was effective for identifying temporal windows containing bird vocalizations and had a relatively low proportion of FNs and FPs.

\subsection{Bird Song Detector vs BirdNET as a Bird Vocalization Detector}

The Bird Song Detector was compared against BirdNET, as a bird vocalization detector, in various configurations. Table \ref{tab:conf_matrix_comparison} presents the confusion matrices for these configurations.

\begin{table*}[!ht]
\footnotesize
\centering
\resizebox{\textwidth}{!}{%
\begin{tabular}{|c|c|c|c|>{\centering\arraybackslash}m{2cm}|>{\centering\arraybackslash}m{2cm}|>{\centering\arraybackslash}m{2cm}|>{\centering\arraybackslash}m{2cm}|}
% \hline
% \multicolumn{8}{|c|}{\textbf{Model evaluation in detection mode: identifying bird sound presence vs. background}} \\
\hline
\multirow{2}{*}{\textbf{Model}}
    & \multirow{2}{*}{\parbox{2.5cm}{\centering \textbf{Fine-tuned}}}
    & \multirow{2}{*}{\parbox{2.5cm}{\centering \textbf{Confidence Score \\ Threshold}}} 
    & \multirow{2}{*}{\parbox{2.5cm}{\centering \textbf{Species \\ List}}} 
    & \multicolumn{2}{c|}{\textbf{Prediction: Bird}} 
    & \multicolumn{2}{c|}{\textbf{Prediction: Background}} \\ \cline{5-8} 
   & & & & {\centering \textbf{GT: Bird \\ (TP)}} 
   & {\centering \textbf{GT: Background \\ (FP)}} 
   & {\centering \textbf{GT: Bird \\ (FN)}} 
   & {\centering \textbf{GT: Background \\ (TN)}} \\ \hline

%\rowcolor{blue!20}
\textbf{\textit{BirdNET}} & \xmark & 0.6 & Full List     & 39  & 0   & 158 & -  \\ \hline
%\rowcolor{blue!20}
\textbf{\textit{BirdNET}} & \xmark & 0.6 & Expert List   & 48  & 1   & 179 & -  \\ \hline
\rowcolor{blue!20}
\textbf{\textit{BirdNET}} & \cmark & 0.6 & Classifier Classes   & 98  & 6   & 211 & -  \\ \hline
\rowcolor{orange!20}
\textbf{\textit{BirdNET}} & \cmark & 0.1 & Classifier Classes  & 245 & 63 & 135 & -   \\ \hline
\textbf{\textit{Bird Song Detector}} & - & 0.15 & - & 196 & 9   & 70  & -   \\ \hline

\rowcolor{blue!20}
\multicolumn{4}{|c|}{\textit{Comparison with 0.6 Confidence Threshold}} & \textit{\textbf{+100\%}} & \textit{+50\%} & \textit{\textbf{-67\%}} & \textit{-} \\ \hline

\rowcolor{orange!20}
\multicolumn{4}{|c|}{\textit{Comparison with 0.1 Confidence Threshold}} & \textit{-20\%} & \textit{\textbf{-86\%}} & \textit{\textbf{-48\%}} & \textit{-} \\ \hline

\end{tabular}
}
\caption{Comparison of confusion matrices for different  models used as \textbf{bird vocalization detectors}, including various configurations of BirdNET (fine-tuned and non fine-tuned, with different confidence thresholds and species lists) and Bird Song Detector model. All models are evaluated solely in terms of detection performance, that is, their ability to distinguish between segments with bird vocalizations and background noise, independently of species classification. The table shows the number of TPs, FPs and FNs for each configuration. The last two rows highlight relative changes in performance when comparing the Bird Song Detector to fine-tuned BirdNET at confidence thresholds of 0.6 (blue) and 0.1 (orange). Bold values indicate improvements made by Bird Song Detector.}
\label{tab:conf_matrix_comparison}
\end{table*}

\subsection{Bird Species Classification}

To assess the impact of the Bird Song Detector on species classification, different classifiers were tested on both raw audio and pre-segmented audio from the detector. Table \ref{tab:classifier_comparison} summarizes the results.

\begin{table*}[!ht]
\scriptsize
\centering
\resizebox{\textwidth}{!}{
\begin{tabular}{|c|c|c|c|c|c|c|c|c|c|}
\hline
\multirow{2}{*}{\textbf{Classifier}} & \multirow{2}{*}{\textbf{Bird Song Detector}} & \multirow{2}{*}{\textbf{Acc.}} & \multicolumn{3}{c|}{\textbf{Macro Avg}} & \multicolumn{3}{c|}{\textbf{Weighted Avg}} & \multirow{2}{*}{\textbf{Idx Pred/Ann}} \\ 
\cline{4-9}
& & & \textbf{Prec.} & \textbf{Rec.} & \textbf{F1} & \textbf{Prec.} & \textbf{Rec.} & \textbf{F1} & \\ 
\hline
\textit{\textbf{BirdNET fine-tuned}} & \xmark  & 0.21  & 0.12  & 0.14  & 0.11  & 0.18  & 0.21  & 0.17  & 1.8046  \\ \hline
\rowcolor{blue!20}
\textit{\textbf{BirdNET fine-tuned}} & \cmark & 0.30 & 0.21 & 0.14 & 0.13 & 0.37 & 0.30 & 0.28 & 0.9183  \\ \hline
\textit{\textbf{Random Forest}} & \xmark  & 0.19  & 0.10  & 0.10  & 0.08  & 0.19  & 0.19  & 0.15  & 0.9059  \\ \hline
\rowcolor{blue!20}
\textit{\textbf{Random Forest}} & \cmark & 0.29 & 0.11 & 0.12 & 0.10 & 0.24 & 0.29 & 0.23 & 0.5435  \\ \hline
\textit{\textbf{ResNet50}} & \xmark  & 0.02  & 0.00  & 0.03  & 0.00  & 0.00  & 0.02  & 0.00  & 3.2682  \\ \hline
\rowcolor{blue!20}
\textit{\textbf{ResNet50}} & \cmark & 0.08 & 0.01 & 0.05 & 0.01 & 0.01 & 0.08 & 0.02 & 0.6306  \\ \hline
\textit{\textbf{MobileNetV2}} & \xmark  & 0.02  & 0.01  & 0.04  & 0.01  & 0.01  & 0.02  & 0.01  & 3.2682  \\ \hline
\rowcolor{blue!20}
\textit{\textbf{MobileNetV2}} & \cmark & 0.08 & 0.01 & 0.04 & 0.01 & 0.02 & 0.08 & 0.02 & 0.6306  \\ \hline
\end{tabular}
}
\caption{Comparison of classifier performance with and without the Bird Song Detector. Shown metrics are Accuracy (Acc.), macro and weighted average (Avg) precision (Prec.), recall (Rec.), F1-score (F1), and an index calculated based on the number of predictions relative to the total number of annotations (Idx Pred/Annn). Shaded rows indicate improvement when using Bird Song Detector. All metrics are better at higher values, except for Idx Pred/Ann, which is optimal when closer to 1.}
\label{tab:classifier_comparison}
\end{table*}

Among the classifiers tested, fine-tuned BirdNET combined with the Bird Song Detector demonstrated the highest performance, achieving an accuracy of 0.30, a weighted F1-score of 0.28, and the best alignment between predictions and ground truth annotations (Idx Pred/Ann = 0.9183). Given its superior performance, a more detailed analysis of this classifier is presented, focusing on per-species performance metrics.

\begin{figure*}[!ht]
    \centering
    \includegraphics[width=1\linewidth]{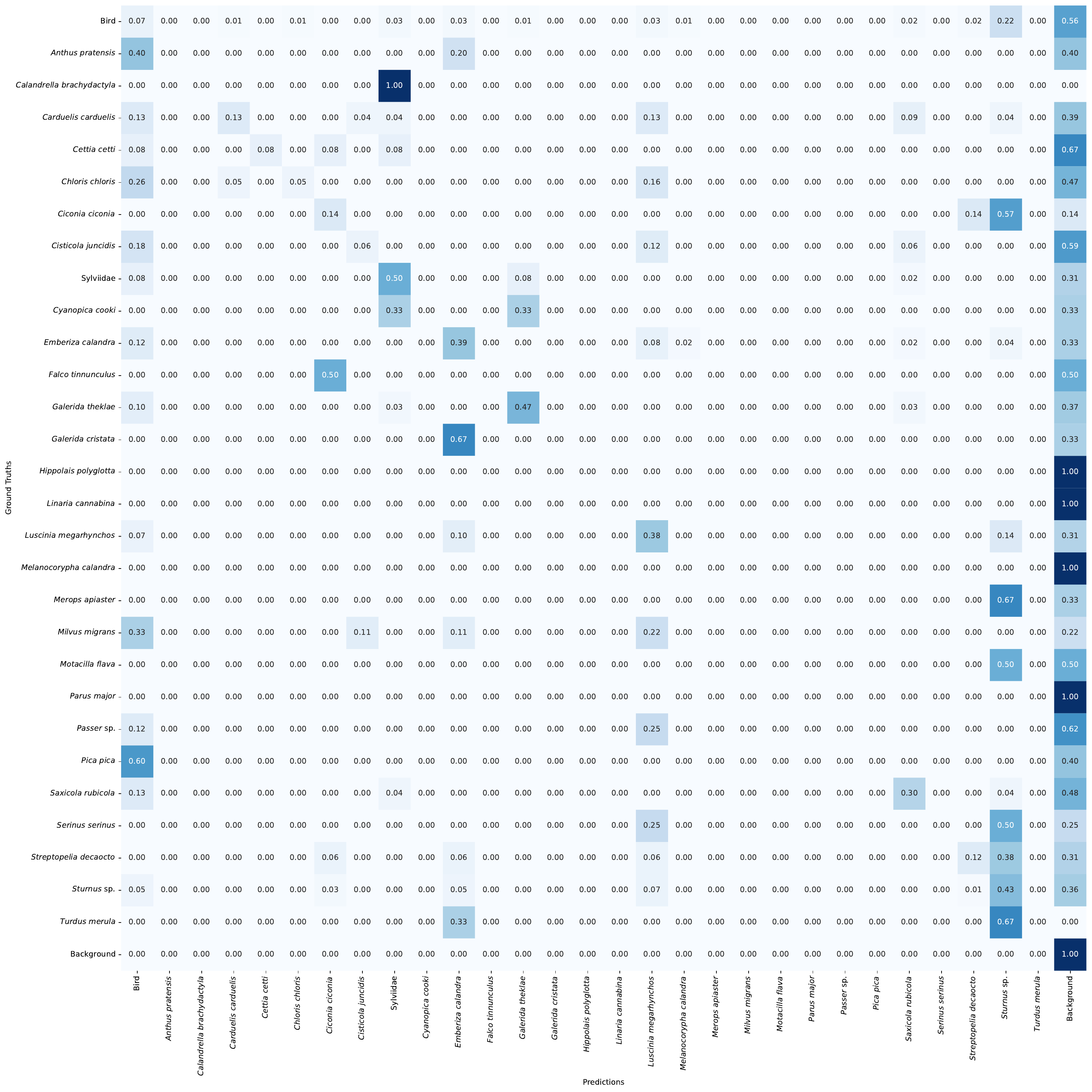}
    \caption{Species-specific normalized confusion matrix showing the performance of the BirdNET classifier on predicted segments from the test dataset, processed by the Bird Song Detector. The values are normalized by rows, with Y-axis labels representing the ground truth species and X-axis labels representing the predicted classes. Darker cells indicate higher correct prediction ratios. The main diagonal (highlighted in bold) corresponds to TP ratios for each species, where predicted labels match the ground truth. Off-diagonal cells represent FPs, where the prediction is incorrect. The ``Bird'' category includes species not present in the classifier’s training data. In such cases, predictions were assigned to the ``Bird'' class, as no better classification could be made.}
    \label{fig:species_confusion_matrix}
\end{figure*}

Figure \ref{fig:species_confusion_matrix} provides the species-specific normalized confusion matrix, offering a visual representation of misclassifications and correct predictions. Darker cells along the diagonal indicate TPs, while off-diagonal cells correspond to FPs and FNs, illustrating common misclassification patterns.

Additionally, to better understand the classification behavior, Table \ref{tab:classification_report} presents the classification report for the Bird Song Detector and fine-tuned BirdNET, showing precision, recall, and F1-score for each species. This allows for an assessment of which species were more accurately identified and which ones posed challenges for classification.

\section{Discussion}
\label{discussion}

% This should explore the significance of the results of the work, not repeat them. A combined Results and Discussion section is often appropriate. Avoid extensive citations and discussion of published literature.

This study presents the methodological development and evaluation of a two-stage pipeline designed to enhance bird species identification in PAM. The results demonstrate that incorporating a Bird Song Detector significantly enhances bird identification in audio recordings. By segmenting recordings and isolating relevant audio regions, the detector effectively reduces FNs while keeping FPs to a minimum. This pre-filtering process ensures that classifiers operate on cleaner, more focused segments, reducing the risk of misclassification due to background noise or overlapping sounds.

% Detectores - Performance of the Bird Song Detector
\noindent\textbf{Evaluation of Detectors. } 
Most publicly available ecoacoustic algorithms \citep{BirdNET-Analyzer, google2024perch} focus primarily on bird species classification rather than the detection of vocalizations. However, general species classification models often struggle in PAM programs deployed in real-world scenarios, as it is already a known case for BirdNET \citep{perez2023birdnet, funosas2024assessing}. Our results confirm that relying solely on BirdNET for binary bird detection leads to high FN and FP rates, which limits its ability to detect bird presence. 

Selecting an appropriate confidence threshold for the Bird Song Detector is key to balance recall and precision. As shown, increasing the threshold reduces FPs but also results in substantial TP losses—up to 99\% at the highest setting. We found that a threshold of 0.15 offered the best compromise, maintaining reasonable recall while limiting the number of non-informative detections processed downstream. This balance is critical in real-world deployments, where excessively low thresholds would flood subsequent stages with noise, while overly conservative thresholds risk missing important vocalizations. The chosen threshold of 0.15 minimizes FN rates without overwhelming the classifier with FP detections, thus optimizing overall system efficiency and ecological reliability.

Fine-tuned BirdNET, when used as a detector with a 0.6 confidence score threshold, failed to detect 68\% of positive samples, meaning a large portion of bird vocalizations were missed. Lowering the confidence threshold to 0.1 improved recall, reducing FNs to 36\%, but at the cost of 20\% FPs, a 950\% rise in FPs. In contrast, the Bird Song Detector achieved a more balanced trade-off, reducing FNs to 26\% while keeping FPs below 5\%. This demonstrates its effectiveness in distinguishing bird vocalizations from background noise while maintaining high precision.

\noindent\textbf{Threshold Optimization and Error Analysis. }
Compared to BirdNET at 0.6 confidence, the Bird Song Detector significantly improved detection performance. Specifically, FNs decreased by 67\% (from 211 to 70), ensuring that far more bird vocalizations were correctly identified. While this came with a 50\% increase in FPs (from 6 to 9), the absolute increase was only three additional FPs. Considering that this evaluation was conducted on the test subset, which comprised approximately 100 to 200 minutes of annotated audio, where the potential for false detections is virtually limitless, an increase of three FPs has a negligible impact on overall performance while higher numbers like 63 would create excessive noise in PAM applications. 

Upon reviewing the Bird Song Detector’s FPs, we found that some errors were due to anthropogenic noises, such as repetitive fence hits, which may share frequency patterns or temporal structures with bird vocalizations. This highlights an ongoing challenge in field recordings, where separating natural and human-made sounds can be difficult. Further refinements, such as incorporating additional non-bird training data or adaptive thresholding strategies, could help reduce these misclassifications.

On the other hand, when compared to BirdNET at the lower 0.1 confidence score threshold, which is BirdNET’s default confidence score threshold setting, the Bird Song Detector further reduced FNs by 48\% and FPs by 86\%, though it resulted in a 20\% decrease in TPs. Despite this slight reduction, the overall improvement in FN minimization and precision suggests that the Bird Song Detector provides a more reliable balance between sensitivity and accuracy. This is particularly relevant in ecological studies, where minimizing FNs is often more critical than reducing FPs, as missed vocalizations can lead to an underestimation of species presence and activity.

Beyond improving bird presence detection, the Bird Song Detector enhances species classification by ensuring that only relevant audio segments are analyzed. Unlike BirdNET, which operates on fixed 3-second windows regardless of whether they contain a vocalization, our detector selects only segments where bird sounds have been detected. This targeted segmentation reduces background noise and mitigates the effects of overlapping calls, two major obstacles in species identification. As a result, downstream classifiers operate with a higher signal-to-noise ratio, leading to greater classification accuracy and a more efficient analysis of bioacoustic data.

% Clasificadores

\noindent\textbf{Comparison of Classifiers and Deep Learning Models. }
The Bird Song Detector significantly improved classification accuracy across all models. Without it, classifiers exhibited higher FP rates and greater misclassification errors, as they processed unfiltered, noisy audio segments. The reduction in Idx Pred/Ann indicates that the Bird Song Detector provided cleaner, more structured data, improving the alignment between predictions and ground truth annotations.

The shown improvements underscore the importance of pre-filtering irrelevant audio segments before classification, as it reduces false detections and improves prediction reliability. The Random Forest classifier also demonstrated notable gains, increasing its accuracy from 0.19 to 0.29 when paired with the Bird Song Detector, suggesting that traditional machine learning models can remain competitive in bioacoustic classification when leveraging BirdNET embeddings \citep{ghani2023global}.

However, despite these improvements, classification performance in this study remained lower than the results reported in the original BirdNET research \citep{kahl2021birdnet}. BirdNET was originally evaluated on clean, single-species recordings, achieving a mean average precision of 0.791, an F0.5 score of 0.414 for annotated soundscapes, and an average correlation of 0.251 with hotspot observations (areas with high species diversity). However, in real-world PAM applications, these conditions do not hold, leading to a notable decline in performance \citep{kahl2021birdnet, perez2023birdnet}. Our findings align with these observations, species classification accuracy remained moderate even after fine-tuning BirdNET and applying the Bird Song Detector. Background noise, species imbalances in training data, and acoustic similarities among certain species contributed to misclassifications, highlighting the ongoing challenges of automated species identification in PAM programs. 

Several species exhibited particularly low precision, recall, and F1-scores, reflecting a failure to correctly classify them. These misclassifications are likely due to class imbalance, species similarity, and dataset limitations. The Bird Song Detector improves classification by filtering out non-bird segments, but it does not eliminate the fundamental challenge of uneven species representation in training data, which remains a limiting factor in overall classification performance.

Despite overall improvements, deep learning models such as ResNet50 and MobileNetV2 performed poorly in both approaches. Their baseline accuracy was extremely low (0.02), and they exhibited high FP rates. Without the Bird Song Detector, these models had very high Idx Pred/Ann values (3.2682), meaning they overpredicted species occurrences far beyond the actual number of vocalizations. Although applying the Bird Song Detector reduced this metric to 0.6306, their classification performance remained inadequate, with weighted F1-scores below 0.02.

These findings suggest that deep learning models pre-trained on image datasets (e.g., \textit{ImageNet}) do not generalize well to bioacoustic classification tasks. Unlike images, bird vocalizations are temporal and frequency-dependent, requiring models to extract patterns across time and frequency domains. The poor performance of \textit{ResNet50} and \textit{MobileNetV2} suggests that models optimized for visual features fail to capture the nuances of bioacoustic signals. Future adaptations for deep learning in this field should explore pretraining on bioacoustic datasets rather than visual datasets like \textit{ImageNet} or using spectrogram-specific architectures designed for temporal pattern recognition \citep{ghani2023global, xiao2022amresnet, xie2022kd, gong2021ast}.

The reduction in Idx Pred/Ann after applying the Bird Song Detector further reinforces the importance of pre-filtering noisy audio segments, even for deep learning architectures. However, the failure of \textit{ResNet50} and \textit{MobileNetV2} to achieve meaningful classification—despite cleaner input data—suggests that architectural modifications, domain-specific pretraining, and tailored feature extraction methods are necessary for deep learning to be effective in species classification.

% Best approeach

\noindent\textbf{Limitations Due to Class Imbalance. }
One limitation of our dataset is the imbalance in the number of annotated vocalizations per species. This imbalance mirrors natural patterns in animal abundance, behavior, and detectability — a well-documented issue in ecological monitoring studies \citep{cui2019class, beery2018recognition, mackenzie2002estimating}. While this poses a challenge for classification performance, it also reflects the operational conditions of real-world PAM applications, reinforcing the need for methods that can handle such skewed data distributions. 

This class imbalance was also evident during dataset splitting and model evaluation. Some species had a disproportionately larger number of training samples than others, while others, such as \textit{Falco tinnunculus} and \textit{Milvus migrans}, were extremely underrepresented. The poor performance of the model was likely due to this imbalance, as classifiers tend to overfit to these dominant classes while performing poorly on underrepresented ones \citep{johnson2019survey, cui2019class, pantazis2024deep}. Although BirdNET’s built-in oversampling was supplemented with \textit{mixup} augmentation, these techniques were not sufficient to counteract imbalance effect—particularly for underrepresented species. Combining diverse augmentation methods has shown promising results in recent work for improving species-level recall in imbalanced datasets \citep{kumar2024improving}.

Prior to this upsampling, 17 species had fewer than 10 instances, and their classification results were generally poor, with most having an F1-score of 0.00. In contrast, only three species had more than 50 instances, and they exhibited relatively higher F1-scores, with \textit{Sturnus} achieving 0.42 and \textit{Emberiza calandra} reaching 0.46. Class frequency and recall appear highly correlated. A clear example is \textit{Anthus pratensis}, which had 42 instances in the test set and achieved a recall of 1.00 but a low precision of 0.16, indicating that the model predicts this class frequently. Conversely, species like \textit{Cettia cetti} and \textit{Galerida cristata} have moderate recall values (0.13 and 0.47, respectively), showing some predictive ability but still suffering from low precision. This suggests that the model frequently predicted this species, even when incorrect, due to its prevalence in the training data. Conversely, species like \textit{Falco tinnunculus} (n=2) or \textit{Milvus migrans} (n=9) had zero recall and precision, indicating that the model was completely unable to recognize them. These bird species normally fly over large altitude and far away from the recorders, so getting more vocalizations that could be labelled and used to train the models would require a more focused sampling.  The model likely overfits to the more frequently occurring classes while underperforms on rare species. This issue is further exacerbated by the presence of species with very small sample sizes, leading to cases where the model never correctly identifies them.

This variation in classification accuracy across species suggests that more balanced training datasets are needed to improve generalization and adaptive thresholding techniques, where confidence scores are dynamically adjusted per species, which could enhance performance \citep{perez2023birdnet, wood2024guidelines}. Despite these challenges, the Bird Song Detector mitigates some imbalanced effects by ensuring classifiers only process relevant bird vocalizations, reducing misclassifications from overwhelming background noise.

The Bird Song Detector helps mitigate data imbalance by ensuring that classifiers only process relevant audio segments of any bird vocalization, reducing misclassification errors caused by excessive background noise. However, even with this improvement, classification performance is still constrained by the lack of diverse and representative training data. Addressing this limitation would require curated datasets with balanced species representation and adaptive thresholding techniques to optimize classification confidence based on species prevalence.

\noindent\textbf{Data Bias in Public Bird Sound Resources. }
Additionally, BirdNET's performance is still influenced by biases in its training data. The public datasets used for BirdNET training (e.g., Xeno-canto \citep{xeno-canto}, Macaulay Library \citep{macaulaylibrary}) are dominated by high-quality focal recordings, which lack overlapping sounds and real-world acoustic complexity. This makes BirdNET less effective in identifying species in dense, multi-species environments, such as Doñana National Park. 

\noindent\textbf{Ecological Implications and Future Applications. }
Although the conservation challenges of Doñana are multiple and require complex solutions from the different social agents involved, we believe that the practical implications of our research for ecological monitoring and conservation of bird populations are significant. Indeed, the integration of Deep Learning models applied to PAM allows for the cost-effective and scalable monitoring of bird diversity, which is essential for tracking bird-species trends over large protected areas. Managers of Doñana can use information from automatic sound classification of bird species to assess the health status of the different habitats present in the area (as many bird species serve as indicators of environmental changes), effectively combining them with direct census monitoring and other methods. This information helps to design conservation measures trying to mitigate bird species diversity loss. While the current work is based on a subset of annotated data, full-scale deployment of this pipeline across the entire dataset from Doñana National Park is ongoing. For this reason, winter migratory species may not be represented in the annotated subset used here. However, future analyses will incorporate these additional recordings. At last, the main aim of the BIRDeep project is to enable a real-time tracking of avian diversity to have accurate information on bird diversity in large, protected areas such as Doñana National Park.

\section{Conclusions}
\label{conclusions}

We have demonstrated the feasibility and advantages of integrating a Bird Song Detector with fine-tuned classifiers for detecting and classifying bird vocalizations in the soundscapes of Doñana National Park. The proposed pipeline enhances both detection and classification accuracy by isolating relevant audio segments before species identification. This approach mitigates many of the challenges faced by traditional classification models, particularly in complex acoustic environments.

The Bird Song Detector plays a crucial role in improving model reliability by significantly reducing FNs while maintaining a manageable FP rate. This ensures that classifiers process only the most relevant segments, leading to more accurate species predictions. However, persistent challenges, such as species misclassification due to class imbalance and the presence of anthropogenic noise, highlight the need for further refinements. Addressing these limitations will require expanding annotated datasets, incorporating adaptive thresholding strategies, and improving classifier fine-tuning for underrepresented species.

Another key challenge lies in the biases present in public-access bioacoustic datasets, which primarily contain clean, focal recordings with minimal background noise. These datasets do not fully represent real-world soundscapes, affecting model generalization in environments like Doñana, where overlapping bird calls and non-bird sounds are common. Future work should focus on developing datasets that better capture the complexity of natural soundscapes and on training models to differentiate between species in multi-species recordings. 

Future research should explore the integration of more advanced bioacoustic deep learning architectures, as models pretrained on image datasets like \textit{ImageNet} may not fully capture the temporal and spectral characteristics of bird vocalizations. Additionally, refining detection models to better distinguish between bird sounds and anthropogenic noise will further enhance the applicability of this approach in real-world conservation settings. 

By providing a structured and scalable method for bird vocalization analysis, the proposed pipeline represents a significant step toward improving automated biodiversity monitoring. With continued advancements in data quality and model adaptation, this framework has the potential to be deployed in diverse ecological contexts, facilitating more precise and efficient monitoring of avian populations and applying bird identification for conservation and management. 

\section*{Data Availability}

The dataset mentioned in this work \citep{alba_marquez_rodriguez_2024} that includes annotated audio, split into train-validation-test and the best data augmentations achieved during experimentation is freely accessible at the following address:  
\url{https://huggingface.co/datasets/GrunCrow/BIRDeep_AudioAnnotations}, {\doi{10.57967/hf/4821}. Metadata follow the Darwin-Core standard \citep{wieczorek2012darwin}.

All code used for this manuscript is freely accessible at the following URL: 
\url{https://github.com/GrunCrow/BIRDeep_BirdSongDetector_NeuralNetworks}. The DOI for this repository is \doi{10.5281/zenodo.14940479}.

The GitHub repository for the Bird Song Detector, which includes an interface, scripts and models to run the detector, is available at:  
\url{https://github.com/GrunCrow/Bird-Song-Detector}. The DOI for this repository is \doi{10.5281/zenodo.15019122}.

\section*{Acknowledgements}

This study has received financial support from the BIRDeep project (TED2021-129871A-I00), which is funded by MICIU/AEI/10.13039/501100011033 and the `European Union NextGenerationEU/PRTR', as well as grants PID2020-115129RJ-I00 from MCIN/AEI/10.13039/501100011033 and PTA2021-020336-I. AMR received a JAE-Intro grant from CSIC (ref. JAEINT23\_EX\_0243). Logistic and technical support for the installation of the recorders and fieldwork assistance were provided by the ICTS-Doñana \citep{ictsdonana}. 

\bibliographystyle{elsarticle-harv} 
\bibliography{main}

\begin{thebibliography}{77}
\expandafter\ifx\csname natexlab\endcsname\relax\def\natexlab#1{#1}\fi
\providecommand{\url}[1]{\texttt{#1}}
\providecommand{\href}[2]{#2}
\providecommand{\path}[1]{#1}
\providecommand{\DOIprefix}{doi:}
\providecommand{\ArXivprefix}{arXiv:}
\providecommand{\URLprefix}{URL: }
\providecommand{\Pubmedprefix}{pmid:}
\providecommand{\doi}[1]{\href{http://dx.doi.org/#1}{\path{#1}}}
\providecommand{\Pubmed}[1]{\href{pmid:#1}{\path{#1}}}
\providecommand{\bibinfo}[2]{#2}
\ifx\xfnm\relax \def\xfnm[#1]{\unskip,\space#1}\fi
%Type = Article
\bibitem[{Audacity(2017)}]{audacity2017audacity}
\bibinfo{author}{Audacity, T.}, \bibinfo{year}{2017}.
\newblock \bibinfo{title}{Audacity}.
\newblock \bibinfo{journal}{The name audacity (R) is a registered trademark of dominic mazzoni retrieved from http://audacity. sourceforge. net} .
%Type = Article
\bibitem[{Beery et~al.(2019)Beery, Morris and Yang}]{Beery_Efficient_Pipeline_for}
\bibinfo{author}{Beery, S.}, \bibinfo{author}{Morris, D.}, \bibinfo{author}{Yang, S.}, \bibinfo{year}{2019}.
\newblock \bibinfo{title}{Efficient pipeline for camera trap image review}.
\newblock \bibinfo{journal}{arXiv preprint arXiv:1907.06772} .
%Type = Inproceedings
\bibitem[{Beery et~al.(2018)Beery, Van~Horn and Perona}]{beery2018recognition}
\bibinfo{author}{Beery, S.}, \bibinfo{author}{Van~Horn, G.}, \bibinfo{author}{Perona, P.}, \bibinfo{year}{2018}.
\newblock \bibinfo{title}{Recognition in terra incognita}, in: \bibinfo{booktitle}{Proceedings of the European conference on computer vision (ECCV)}, pp. \bibinfo{pages}{456--473}.
%Type = Misc
\bibitem[{{Birdeep.org}(2025)}]{birdeeporg}
\bibinfo{author}{{Birdeep.org}}, \bibinfo{year}{2025}.
\newblock \bibinfo{title}{{Birdeep}}.
\newblock \bibinfo{howpublished}{\url{https://birdeeporg.github.io}}.
\newblock \bibinfo{note}{Accessed: 2025-03-13}.
%Type = Article
\bibitem[{Breiman(2001)}]{breiman2001random}
\bibinfo{author}{Breiman, L.}, \bibinfo{year}{2001}.
\newblock \bibinfo{title}{Random forests}.
\newblock \bibinfo{journal}{Machine learning} \bibinfo{volume}{45}, \bibinfo{pages}{5--32}.
%Type = Article
\bibitem[{Brunk et~al.(2023)Brunk, Guti{\'e}rrez, Peery, Cansler, Kahl and Wood}]{brunk2023quail}
\bibinfo{author}{Brunk, K.M.}, \bibinfo{author}{Guti{\'e}rrez, R.}, \bibinfo{author}{Peery, M.Z.}, \bibinfo{author}{Cansler, C.A.}, \bibinfo{author}{Kahl, S.}, \bibinfo{author}{Wood, C.M.}, \bibinfo{year}{2023}.
\newblock \bibinfo{title}{Quail on fire: changing fire regimes may benefit mountain quail in fire-adapted forests}.
\newblock \bibinfo{journal}{Fire Ecology} \bibinfo{volume}{19}, \bibinfo{pages}{1--13}.
%Type = Article
\bibitem[{Camacho et~al.(2022)Camacho, Negro, Elmberg, Fox, Nagy, Pain and Green}]{camacho2022groundwater}
\bibinfo{author}{Camacho, C.}, \bibinfo{author}{Negro, J.J.}, \bibinfo{author}{Elmberg, J.}, \bibinfo{author}{Fox, A.D.}, \bibinfo{author}{Nagy, S.}, \bibinfo{author}{Pain, D.J.}, \bibinfo{author}{Green, A.J.}, \bibinfo{year}{2022}.
\newblock \bibinfo{title}{Groundwater extraction poses extreme threat to do{\~n}ana world heritage site}.
\newblock \bibinfo{journal}{Nature Ecology \& Evolution} \bibinfo{volume}{6}, \bibinfo{pages}{654--655}.
%Type = Article
\bibitem[{Campo-Celada et~al.(2022)Campo-Celada, Jordano, Ben{\'\i}tez-L{\'o}pez, Guti{\'e}rrez-Exp{\'o}sito, Rabad{\'a}n-Gonz{\'a}lez and Mendoza}]{campo2022assessing}
\bibinfo{author}{Campo-Celada, M.}, \bibinfo{author}{Jordano, P.}, \bibinfo{author}{Ben{\'\i}tez-L{\'o}pez, A.}, \bibinfo{author}{Guti{\'e}rrez-Exp{\'o}sito, C.}, \bibinfo{author}{Rabad{\'a}n-Gonz{\'a}lez, J.}, \bibinfo{author}{Mendoza, I.}, \bibinfo{year}{2022}.
\newblock \bibinfo{title}{Assessing short and long-term variations in diversity, timing and body condition of frugivorous birds}.
\newblock \bibinfo{journal}{Oikos} \bibinfo{volume}{2022}.
%Type = Article
\bibitem[{Carvalho and Gomes(2023)}]{carvalho2023automatic}
\bibinfo{author}{Carvalho, S.}, \bibinfo{author}{Gomes, E.F.}, \bibinfo{year}{2023}.
\newblock \bibinfo{title}{Automatic classification of bird sounds: using mfcc and mel spectrogram features with deep learning}.
\newblock \bibinfo{journal}{Vietnam Journal of Computer Science} \bibinfo{volume}{10}, \bibinfo{pages}{39--54}.
%Type = Misc
\bibitem[{Chollet et~al.(2015)}]{chollet2015keras}
\bibinfo{author}{Chollet, F.}, et~al., \bibinfo{year}{2015}.
\newblock \bibinfo{title}{Keras}.
\newblock \URLprefix \url{https://github.com/fchollet/keras}.
%Type = Article
\bibitem[{Clark et~al.(2023)Clark, Salas, Baligar, Quinn, Snyder, Leland, Schackwitz, Goetz and Newsam}]{clark2023effect}
\bibinfo{author}{Clark, M.L.}, \bibinfo{author}{Salas, L.}, \bibinfo{author}{Baligar, S.}, \bibinfo{author}{Quinn, C.A.}, \bibinfo{author}{Snyder, R.L.}, \bibinfo{author}{Leland, D.}, \bibinfo{author}{Schackwitz, W.}, \bibinfo{author}{Goetz, S.J.}, \bibinfo{author}{Newsam, S.}, \bibinfo{year}{2023}.
\newblock \bibinfo{title}{The effect of soundscape composition on bird vocalization classification in a citizen science biodiversity monitoring project}.
\newblock \bibinfo{journal}{Ecological Informatics} \bibinfo{volume}{75}, \bibinfo{pages}{102065}.
%Type = Inproceedings
\bibitem[{Cui et~al.(2019)Cui, Jia, Lin, Song and Belongie}]{cui2019class}
\bibinfo{author}{Cui, Y.}, \bibinfo{author}{Jia, M.}, \bibinfo{author}{Lin, T.Y.}, \bibinfo{author}{Song, Y.}, \bibinfo{author}{Belongie, S.}, \bibinfo{year}{2019}.
\newblock \bibinfo{title}{Class-balanced loss based on effective number of samples}, in: \bibinfo{booktitle}{Proceedings of the IEEE/CVF conference on computer vision and pattern recognition}, pp. \bibinfo{pages}{9268--9277}.
%Type = Article
\bibitem[{Darras et~al.(2019)Darras, Bat{\'a}ry, Furnas, Grass, Mulyani and Tscharntke}]{darras2019autonomous}
\bibinfo{author}{Darras, K.}, \bibinfo{author}{Bat{\'a}ry, P.}, \bibinfo{author}{Furnas, B.J.}, \bibinfo{author}{Grass, I.}, \bibinfo{author}{Mulyani, Y.A.}, \bibinfo{author}{Tscharntke, T.}, \bibinfo{year}{2019}.
\newblock \bibinfo{title}{Autonomous sound recording outperforms human observation for sampling birds: a systematic map and user guide}.
\newblock \bibinfo{journal}{Ecological Applications} \bibinfo{volume}{29}, \bibinfo{pages}{e01954}.
%Type = Article
\bibitem[{Farley et~al.(2018)Farley, Dawson, Goring and Williams}]{farley2018situating}
\bibinfo{author}{Farley, S.S.}, \bibinfo{author}{Dawson, A.}, \bibinfo{author}{Goring, S.J.}, \bibinfo{author}{Williams, J.W.}, \bibinfo{year}{2018}.
\newblock \bibinfo{title}{Situating ecology as a big-data science: current advances, challenges, and solutions}.
\newblock \bibinfo{journal}{BioScience} \bibinfo{volume}{68}, \bibinfo{pages}{563--576}.
%Type = Misc
\bibitem[{canto Foundation(2024)}]{xeno-canto}
\bibinfo{author}{canto Foundation, X.}, \bibinfo{year}{2024}.
\newblock \bibinfo{title}{Xeno-canto: Bird sounds from around the world}.
\newblock \URLprefix \url{https://www.xeno-canto.org}. \bibinfo{note}{accessed: 2024-07-13}.
%Type = Article
\bibitem[{Funosas et~al.(2024)Funosas, Barbaro, Schill{\'e}, Elger, Castagneyrol and Cauchoix}]{funosas2024assessing}
\bibinfo{author}{Funosas, D.}, \bibinfo{author}{Barbaro, L.}, \bibinfo{author}{Schill{\'e}, L.}, \bibinfo{author}{Elger, A.}, \bibinfo{author}{Castagneyrol, B.}, \bibinfo{author}{Cauchoix, M.}, \bibinfo{year}{2024}.
\newblock \bibinfo{title}{Assessing the potential of birdnet to infer european bird communities from large-scale ecoacoustic data}.
\newblock \bibinfo{journal}{Ecological Indicators} \bibinfo{volume}{164}, \bibinfo{pages}{112146}.
%Type = Misc
\bibitem[{Garc{\'\i}a et~al.(2000)Garc{\'\i}a, Ib{\'a}{\~n}ez, Garrido, Arroyo, M{\'a}{\~n}ez and Calder{\'o}n}]{garcia2000prontuario}
\bibinfo{author}{Garc{\'\i}a, L.}, \bibinfo{author}{Ib{\'a}{\~n}ez, F.}, \bibinfo{author}{Garrido, H.}, \bibinfo{author}{Arroyo, J.}, \bibinfo{author}{M{\'a}{\~n}ez, M.}, \bibinfo{author}{Calder{\'o}n, J.}, \bibinfo{year}{2000}.
\newblock \bibinfo{title}{Prontuario de las aves de do{\~n}ana. anuario ornitol{\'o}gico de do{\~n}ana, n{\textordmasculine} 0, diciembre 2000}.
%Type = Article
\bibitem[{Garrido et~al.(2004)Garrido, Arroyo, Garc{\'\i}a, Ib{\'a}{\~n}ez, M{\'a}{\~n}ez and V{\'a}zquez}]{garrido2004anuario}
\bibinfo{author}{Garrido, H.}, \bibinfo{author}{Arroyo, J.}, \bibinfo{author}{Garc{\'\i}a, L.}, \bibinfo{author}{Ib{\'a}{\~n}ez, F.}, \bibinfo{author}{M{\'a}{\~n}ez, M.}, \bibinfo{author}{V{\'a}zquez, M.}, \bibinfo{year}{2004}.
\newblock \bibinfo{title}{Anuario ornitol{\'o}gico de do{\~n}ana, n{\textordmasculine} 1 (septiembre 1999--agosto 2001)}.
\newblock \bibinfo{journal}{Estaci{\'o}n Biol{\'o}gica de Do{\~n}ana y Ayuntamiento de Almonte, Almonte (Huelva)(in Spanish)} .
%Type = Article
\bibitem[{Ghani et~al.(2023)Ghani, Denton, Kahl and Klinck}]{ghani2023global}
\bibinfo{author}{Ghani, B.}, \bibinfo{author}{Denton, T.}, \bibinfo{author}{Kahl, S.}, \bibinfo{author}{Klinck, H.}, \bibinfo{year}{2023}.
\newblock \bibinfo{title}{Global birdsong embeddings enable superior transfer learning for bioacoustic classification}.
\newblock \bibinfo{journal}{Scientific Reports} \bibinfo{volume}{13}, \bibinfo{pages}{22876}.
%Type = Article
\bibitem[{Ghani et~al.(2024)Ghani, Kalkman, Planqu{\'e}, Vellinga, Gill and Stowell}]{ghani2024generalization}
\bibinfo{author}{Ghani, B.}, \bibinfo{author}{Kalkman, V.J.}, \bibinfo{author}{Planqu{\'e}, B.}, \bibinfo{author}{Vellinga, W.P.}, \bibinfo{author}{Gill, L.}, \bibinfo{author}{Stowell, D.}, \bibinfo{year}{2024}.
\newblock \bibinfo{title}{Generalization in birdsong classification: impact of transfer learning methods and dataset characteristics}.
\newblock \bibinfo{journal}{arXiv preprint arXiv:2409.15383} .
%Type = Article
\bibitem[{Gibb et~al.(2019)Gibb, Browning, Glover-Kapfer and Jones}]{gibb2019emerging}
\bibinfo{author}{Gibb, R.}, \bibinfo{author}{Browning, E.}, \bibinfo{author}{Glover-Kapfer, P.}, \bibinfo{author}{Jones, K.E.}, \bibinfo{year}{2019}.
\newblock \bibinfo{title}{Emerging opportunities and challenges for passive acoustics in ecological assessment and monitoring}.
\newblock \bibinfo{journal}{Methods in Ecology and Evolution} \bibinfo{volume}{10}, \bibinfo{pages}{169--185}.
%Type = Article
\bibitem[{Gong et~al.(2021)Gong, Chung and Glass}]{gong2021ast}
\bibinfo{author}{Gong, Y.}, \bibinfo{author}{Chung, Y.A.}, \bibinfo{author}{Glass, J.}, \bibinfo{year}{2021}.
\newblock \bibinfo{title}{Ast: Audio spectrogram transformer}.
\newblock \bibinfo{journal}{arXiv preprint arXiv:2104.01778} .
%Type = Misc
\bibitem[{Goodfellow(2016)}]{goodfellow2016deep}
\bibinfo{author}{Goodfellow, I.}, \bibinfo{year}{2016}.
\newblock \bibinfo{title}{Deep learning}.
%Type = Misc
\bibitem[{{Google Research}(2024)}]{google2024perch}
\bibinfo{author}{{Google Research}}, \bibinfo{year}{2024}.
\newblock \bibinfo{title}{Perch: A bioacoustics research project}.
\newblock \URLprefix \url{https://github.com/google-research/perch}. \bibinfo{note}{accessed: 2024-07-15}.
%Type = Article
\bibitem[{Green et~al.(2016)Green, Bustamante, Janss, Fern{\'a}ndez-Zamudio, D{\'\i}az-Paniagua et~al.}]{green2016donana}
\bibinfo{author}{Green, A.J.}, \bibinfo{author}{Bustamante, J.}, \bibinfo{author}{Janss, G.}, \bibinfo{author}{Fern{\'a}ndez-Zamudio, R.}, \bibinfo{author}{D{\'\i}az-Paniagua, C.}, et~al., \bibinfo{year}{2016}.
\newblock \bibinfo{title}{Do{\~n}ana wetlands} .
%Type = Article
\bibitem[{Green et~al.(2024)Green, Guardiola-Albert, Bravo-Utrera, Bustamante, Camacho, Camacho, Contreras-Arribas, Espinar, Gil-Gil, Gomez-Mestre et~al.}]{green2024groundwater}
\bibinfo{author}{Green, A.J.}, \bibinfo{author}{Guardiola-Albert, C.}, \bibinfo{author}{Bravo-Utrera, M.{\'A}.}, \bibinfo{author}{Bustamante, J.}, \bibinfo{author}{Camacho, A.}, \bibinfo{author}{Camacho, C.}, \bibinfo{author}{Contreras-Arribas, E.}, \bibinfo{author}{Espinar, J.L.}, \bibinfo{author}{Gil-Gil, T.}, \bibinfo{author}{Gomez-Mestre, I.}, et~al., \bibinfo{year}{2024}.
\newblock \bibinfo{title}{Groundwater abstraction has caused extensive ecological damage to the do{\~n}ana world heritage site, spain}.
\newblock \bibinfo{journal}{Wetlands} \bibinfo{volume}{44}, \bibinfo{pages}{20}.
%Type = Article
\bibitem[{Gregory and van Strien(2010)}]{gregory2010wild}
\bibinfo{author}{Gregory, R.D.}, \bibinfo{author}{van Strien, A.}, \bibinfo{year}{2010}.
\newblock \bibinfo{title}{Wild bird indicators: using composite population trends of birds as measures of environmental health}.
\newblock \bibinfo{journal}{Ornithological Science} \bibinfo{volume}{9}, \bibinfo{pages}{3--22}.
%Type = Article
\bibitem[{Hamer et~al.(2023)Hamer, Triantafillou, van Merrienboer, Kahl, Klinck, Denton and Dumoulin}]{hamer2023birb}
\bibinfo{author}{Hamer, J.}, \bibinfo{author}{Triantafillou, E.}, \bibinfo{author}{van Merrienboer, B.}, \bibinfo{author}{Kahl, S.}, \bibinfo{author}{Klinck, H.}, \bibinfo{author}{Denton, T.}, \bibinfo{author}{Dumoulin, V.}, \bibinfo{year}{2023}.
\newblock \bibinfo{title}{Birb: A generalization benchmark for information retrieval in bioacoustics}.
\newblock \bibinfo{journal}{arXiv preprint arXiv:2312.07439} .
%Type = Article
\bibitem[{Hardy(2010)}]{hardy2010pareto}
\bibinfo{author}{Hardy, M.}, \bibinfo{year}{2010}.
\newblock \bibinfo{title}{Pareto’s law}.
\newblock \bibinfo{journal}{The Mathematical Intelligencer} \bibinfo{volume}{32}, \bibinfo{pages}{38--43}.
%Type = Article
\bibitem[{Harris et~al.(2020)Harris, Millman, van~der Walt, Gommers, Virtanen, Cournapeau, Wieser, Taylor, Berg, Smith, Kern, Picus, Hoyer, van Kerkwijk, Brett, Haldane, del R{\'{i}}o, Wiebe, Peterson, G{\'{e}}rard-Marchant, Sheppard, Reddy, Weckesser, Abbasi, Gohlke and Oliphant}]{harris2020array}
\bibinfo{author}{Harris, C.R.}, \bibinfo{author}{Millman, K.J.}, \bibinfo{author}{van~der Walt, S.J.}, \bibinfo{author}{Gommers, R.}, \bibinfo{author}{Virtanen, P.}, \bibinfo{author}{Cournapeau, D.}, \bibinfo{author}{Wieser, E.}, \bibinfo{author}{Taylor, J.}, \bibinfo{author}{Berg, S.}, \bibinfo{author}{Smith, N.J.}, \bibinfo{author}{Kern, R.}, \bibinfo{author}{Picus, M.}, \bibinfo{author}{Hoyer, S.}, \bibinfo{author}{van Kerkwijk, M.H.}, \bibinfo{author}{Brett, M.}, \bibinfo{author}{Haldane, A.}, \bibinfo{author}{del R{\'{i}}o, J.F.}, \bibinfo{author}{Wiebe, M.}, \bibinfo{author}{Peterson, P.}, \bibinfo{author}{G{\'{e}}rard-Marchant, P.}, \bibinfo{author}{Sheppard, K.}, \bibinfo{author}{Reddy, T.}, \bibinfo{author}{Weckesser, W.}, \bibinfo{author}{Abbasi, H.}, \bibinfo{author}{Gohlke, C.}, \bibinfo{author}{Oliphant, T.E.}, \bibinfo{year}{2020}.
\newblock \bibinfo{title}{Array programming with {NumPy}}.
\newblock \bibinfo{journal}{Nature} \bibinfo{volume}{585}, \bibinfo{pages}{357--362}.
\newblock \URLprefix \url{https://doi.org/10.1038/s41586-020-2649-2}, \DOIprefix\doi{10.1038/s41586-020-2649-2}.
%Type = Book
\bibitem[{Hastie et~al.(2009)Hastie, Tibshirani, Friedman and Friedman}]{hastie2009elements}
\bibinfo{author}{Hastie, T.}, \bibinfo{author}{Tibshirani, R.}, \bibinfo{author}{Friedman, J.H.}, \bibinfo{author}{Friedman, J.H.}, \bibinfo{year}{2009}.
\newblock \bibinfo{title}{The elements of statistical learning: data mining, inference, and prediction}. volume~\bibinfo{volume}{2}.
\newblock \bibinfo{publisher}{Springer}.
%Type = Article
\bibitem[{Hill et~al.(2018)Hill, Prince, Pi{\~n}a~Covarrubias, Doncaster, Snaddon and Rogers}]{hill2018audiomoth}
\bibinfo{author}{Hill, A.P.}, \bibinfo{author}{Prince, P.}, \bibinfo{author}{Pi{\~n}a~Covarrubias, E.}, \bibinfo{author}{Doncaster, C.P.}, \bibinfo{author}{Snaddon, J.L.}, \bibinfo{author}{Rogers, A.}, \bibinfo{year}{2018}.
\newblock \bibinfo{title}{Audiomoth: Evaluation of a smart open acoustic device for monitoring biodiversity and the environment}.
\newblock \bibinfo{journal}{Methods in Ecology and Evolution} \bibinfo{volume}{9}, \bibinfo{pages}{1199--1211}.
%Type = Article
\bibitem[{Hill et~al.(2019)Hill, Prince, Snaddon, Doncaster and Rogers}]{hill2019audiomoth}
\bibinfo{author}{Hill, A.P.}, \bibinfo{author}{Prince, P.}, \bibinfo{author}{Snaddon, J.L.}, \bibinfo{author}{Doncaster, C.P.}, \bibinfo{author}{Rogers, A.}, \bibinfo{year}{2019}.
\newblock \bibinfo{title}{Audiomoth: A low-cost acoustic device for monitoring biodiversity and the environment}.
\newblock \bibinfo{journal}{HardwareX} \bibinfo{volume}{6}, \bibinfo{pages}{e00073}.
%Type = Misc
\bibitem[{{ICTS Doñana}(2025)}]{ictsdonana}
\bibinfo{author}{{ICTS Doñana}}, \bibinfo{year}{2025}.
\newblock \bibinfo{title}{Icts doñana}.
\newblock \URLprefix \url{https://icts-donana.csic.es/}. \bibinfo{note}{accessed: 2025-03-13}.
%Type = Misc
\bibitem[{Jocher et~al.(2023)Jocher, Chaurasia and Qiu}]{Jocher_Ultralytics_YOLO_2023}
\bibinfo{author}{Jocher, G.}, \bibinfo{author}{Chaurasia, A.}, \bibinfo{author}{Qiu, J.}, \bibinfo{year}{2023}.
\newblock \bibinfo{title}{{Ultralytics YOLO}}.
\newblock \URLprefix \url{https://github.com/ultralytics/ultralytics}.
%Type = Article
\bibitem[{Johnson and Khoshgoftaar(2019)}]{johnson2019survey}
\bibinfo{author}{Johnson, J.M.}, \bibinfo{author}{Khoshgoftaar, T.M.}, \bibinfo{year}{2019}.
\newblock \bibinfo{title}{Survey on deep learning with class imbalance}.
\newblock \bibinfo{journal}{Journal of big data} \bibinfo{volume}{6}, \bibinfo{pages}{1--54}.
%Type = Misc
\bibitem[{Kahl(2021)}]{BirdNET-Analyzer}
\bibinfo{author}{Kahl, S.}, \bibinfo{year}{2021}.
\newblock \bibinfo{title}{Birdnet-analyzer}.
\newblock \bibinfo{howpublished}{\url{https://github.com/kahst/BirdNET-Analyzer}}.
%Type = Article
\bibitem[{Kahl et~al.(2021)Kahl, Wood, Eibl and Klinck}]{kahl2021birdnet}
\bibinfo{author}{Kahl, S.}, \bibinfo{author}{Wood, C.M.}, \bibinfo{author}{Eibl, M.}, \bibinfo{author}{Klinck, H.}, \bibinfo{year}{2021}.
\newblock \bibinfo{title}{Birdnet: A deep learning solution for avian diversity monitoring}.
\newblock \bibinfo{journal}{Ecological Informatics} \bibinfo{volume}{61}, \bibinfo{pages}{101236}.
%Type = Article
\bibitem[{Kattenborn et~al.(2022)Kattenborn, Schiefer, Frey, Feilhauer, Mahecha and Dormann}]{kattenborn2022spatially}
\bibinfo{author}{Kattenborn, T.}, \bibinfo{author}{Schiefer, F.}, \bibinfo{author}{Frey, J.}, \bibinfo{author}{Feilhauer, H.}, \bibinfo{author}{Mahecha, M.D.}, \bibinfo{author}{Dormann, C.F.}, \bibinfo{year}{2022}.
\newblock \bibinfo{title}{Spatially autocorrelated training and validation samples inflate performance assessment of convolutional neural networks}.
\newblock \bibinfo{journal}{ISPRS Open Journal of Photogrammetry and Remote Sensing} \bibinfo{volume}{5}, \bibinfo{pages}{100018}.
%Type = Article
\bibitem[{Keen et~al.(2021)Keen, Odom, Webster, Kohn, Wright and Araya-Salas}]{keen2021machine}
\bibinfo{author}{Keen, S.C.}, \bibinfo{author}{Odom, K.J.}, \bibinfo{author}{Webster, M.S.}, \bibinfo{author}{Kohn, G.M.}, \bibinfo{author}{Wright, T.F.}, \bibinfo{author}{Araya-Salas, M.}, \bibinfo{year}{2021}.
\newblock \bibinfo{title}{A machine learning approach for classifying and quantifying acoustic diversity}.
\newblock \bibinfo{journal}{Methods in ecology and evolution} \bibinfo{volume}{12}, \bibinfo{pages}{1213--1225}.
%Type = Article
\bibitem[{Kumar et~al.(2024)Kumar, Schlosser, Kahl and Kowerko}]{kumar2024improving}
\bibinfo{author}{Kumar, A.S.}, \bibinfo{author}{Schlosser, T.}, \bibinfo{author}{Kahl, S.}, \bibinfo{author}{Kowerko, D.}, \bibinfo{year}{2024}.
\newblock \bibinfo{title}{Improving learning-based birdsong classification by utilizing combined audio augmentation strategies}.
\newblock \bibinfo{journal}{Ecological Informatics} \bibinfo{volume}{82}, \bibinfo{pages}{102699}.
%Type = Article
\bibitem[{Lalor et~al.(2017)Lalor, Wu and Yu}]{improvingmachinelearning_lalor}
\bibinfo{author}{Lalor, J.P.}, \bibinfo{author}{Wu, H.}, \bibinfo{author}{Yu, H.}, \bibinfo{year}{2017}.
\newblock \bibinfo{title}{Improving machine learning ability with finetuning}.
\newblock \bibinfo{journal}{CoRR, abs/1702.08563. Version} \bibinfo{volume}{1}.
%Type = Article
\bibitem[{Lauha et~al.(2022)Lauha, Somervuo, Lehikoinen, Geres, Richter, Seibold and Ovaskainen}]{lauha2022domain}
\bibinfo{author}{Lauha, P.}, \bibinfo{author}{Somervuo, P.}, \bibinfo{author}{Lehikoinen, P.}, \bibinfo{author}{Geres, L.}, \bibinfo{author}{Richter, T.}, \bibinfo{author}{Seibold, S.}, \bibinfo{author}{Ovaskainen, O.}, \bibinfo{year}{2022}.
\newblock \bibinfo{title}{Domain-specific neural networks improve automated bird sound recognition already with small amount of local data}.
\newblock \bibinfo{journal}{Methods in Ecology and Evolution} \bibinfo{volume}{13}, \bibinfo{pages}{2799--2810}.
%Type = Misc
\bibitem[{{Macaulay Library}(2024)}]{macaulaylibrary}
\bibinfo{author}{{Macaulay Library}}, \bibinfo{year}{2024}.
\newblock \bibinfo{title}{The world's premier scientific archive of natural history audio, video, and photographs}.
\newblock \bibinfo{howpublished}{\url{https://www.macaulaylibrary.org/about/history/}}.
\newblock \bibinfo{note}{Accessed: 2024-07-13}.
%Type = Article
\bibitem[{MacKenzie et~al.(2002)MacKenzie, Nichols, Lachman, Droege, Andrew~Royle and Langtimm}]{mackenzie2002estimating}
\bibinfo{author}{MacKenzie, D.I.}, \bibinfo{author}{Nichols, J.D.}, \bibinfo{author}{Lachman, G.B.}, \bibinfo{author}{Droege, S.}, \bibinfo{author}{Andrew~Royle, J.}, \bibinfo{author}{Langtimm, C.A.}, \bibinfo{year}{2002}.
\newblock \bibinfo{title}{Estimating site occupancy rates when detection probabilities are less than one}.
\newblock \bibinfo{journal}{Ecology} \bibinfo{volume}{83}, \bibinfo{pages}{2248--2255}.
%Type = Misc
\bibitem[{M{\'a}rquez-Rodr{\'i}guez et~al.(2025)M{\'a}rquez-Rodr{\'i}guez, Muñoz-Mohedano, Mar{\'i}n-Jim{\'e}nez, Santamar{\'i}a-Garc{\'i}a, Bastianelli and Mendoza}]{alba_marquez_rodriguez_2024}
\bibinfo{author}{M{\'a}rquez-Rodr{\'i}guez, A.}, \bibinfo{author}{Muñoz-Mohedano, M.{\'A}.}, \bibinfo{author}{Mar{\'i}n-Jim{\'e}nez, M.J.}, \bibinfo{author}{Santamar{\'i}a-Garc{\'i}a, E.}, \bibinfo{author}{Bastianelli, G.}, \bibinfo{author}{Mendoza, I.}, \bibinfo{year}{2025}.
\newblock \bibinfo{title}{Birdeepaudioannotations (revision 4cf0456)}.
\newblock \URLprefix \url{https://huggingface.co/datasets/GrunCrow/BIRDeep_AudioAnnotations}, \DOIprefix\doi{10.57967/hf/4821}.
%Type = Article
\bibitem[{Martin et~al.(2022)Martin, Adam, Obin and Dufour}]{martin2022rookognise}
\bibinfo{author}{Martin, K.}, \bibinfo{author}{Adam, O.}, \bibinfo{author}{Obin, N.}, \bibinfo{author}{Dufour, V.}, \bibinfo{year}{2022}.
\newblock \bibinfo{title}{Rookognise: Acoustic detection and identification of individual rooks in field recordings using multi-task neural networks}.
\newblock \bibinfo{journal}{Ecological Informatics} \bibinfo{volume}{72}, \bibinfo{pages}{101818}.
%Type = Inproceedings
\bibitem[{McFee et~al.(2015)McFee, Raffel, Liang, Ellis, McVicar, Battenberg and Nieto}]{librosa2025}
\bibinfo{author}{McFee, B.}, \bibinfo{author}{Raffel, C.}, \bibinfo{author}{Liang, D.}, \bibinfo{author}{Ellis, D.P.}, \bibinfo{author}{McVicar, M.}, \bibinfo{author}{Battenberg, E.}, \bibinfo{author}{Nieto, O.}, \bibinfo{year}{2015}.
\newblock \bibinfo{title}{librosa: Audio and music signal analysis in python}, in: \bibinfo{booktitle}{Proceedings of the 14th python in science conference}.
%Type = Article
\bibitem[{Metcalf et~al.(2023)Metcalf, Abrahams, Ashington, Baker, Bradfer-Lawrence, Browning, Carruthers-Jones, Darby, Dick, Eldridge et~al.}]{metcalf2023good}
\bibinfo{author}{Metcalf, O.}, \bibinfo{author}{Abrahams, C.}, \bibinfo{author}{Ashington, B.}, \bibinfo{author}{Baker, E.}, \bibinfo{author}{Bradfer-Lawrence, T.}, \bibinfo{author}{Browning, E.}, \bibinfo{author}{Carruthers-Jones, J.}, \bibinfo{author}{Darby, J.}, \bibinfo{author}{Dick, J.}, \bibinfo{author}{Eldridge, A.}, et~al., \bibinfo{year}{2023}.
\newblock \bibinfo{title}{Good practice guidelines for long-term ecoacoustic monitoring in the uk} .
%Type = Book
\bibitem[{Murphy(2012)}]{murphy2012machine}
\bibinfo{author}{Murphy, K.P.}, \bibinfo{year}{2012}.
\newblock \bibinfo{title}{Machine learning: a probabilistic perspective}.
\newblock \bibinfo{publisher}{MIT press}.
%Type = Article
\bibitem[{Padilla et~al.(2021)Padilla, Passos, Dias, Netto and Da~Silva}]{padilla2021comparative}
\bibinfo{author}{Padilla, R.}, \bibinfo{author}{Passos, W.L.}, \bibinfo{author}{Dias, T.L.}, \bibinfo{author}{Netto, S.L.}, \bibinfo{author}{Da~Silva, E.A.}, \bibinfo{year}{2021}.
\newblock \bibinfo{title}{A comparative analysis of object detection metrics with a companion open-source toolkit}.
\newblock \bibinfo{journal}{Electronics} \bibinfo{volume}{10}, \bibinfo{pages}{279}.
%Type = Article
\bibitem[{Pantazis et~al.(2024)Pantazis, Bevan, Pringle, Ferreira, Ingram, Madsen, Thomas, Thanet, Silwal, Rayamajhi et~al.}]{pantazis2024deep}
\bibinfo{author}{Pantazis, O.}, \bibinfo{author}{Bevan, P.}, \bibinfo{author}{Pringle, H.}, \bibinfo{author}{Ferreira, G.B.}, \bibinfo{author}{Ingram, D.J.}, \bibinfo{author}{Madsen, E.}, \bibinfo{author}{Thomas, L.}, \bibinfo{author}{Thanet, D.R.}, \bibinfo{author}{Silwal, T.}, \bibinfo{author}{Rayamajhi, S.}, et~al., \bibinfo{year}{2024}.
\newblock \bibinfo{title}{Deep learning-based ecological analysis of camera trap images is impacted by training data quality and size}.
\newblock \bibinfo{journal}{arXiv preprint arXiv:2408.14348} .
%Type = Inproceedings
\bibitem[{Pasupa and Sunhem(2016)}]{pasupa2016comparison}
\bibinfo{author}{Pasupa, K.}, \bibinfo{author}{Sunhem, W.}, \bibinfo{year}{2016}.
\newblock \bibinfo{title}{A comparison between shallow and deep architecture classifiers on small dataset}, in: \bibinfo{booktitle}{2016 8th International Conference on Information Technology and Electrical Engineering (ICITEE)}, \bibinfo{organization}{IEEE}. pp. \bibinfo{pages}{1--6}.
%Type = Article
\bibitem[{Pedregosa et~al.(2011)Pedregosa, Varoquaux, Gramfort, Michel, Thirion, Grisel, Blondel, Prettenhofer, Weiss, Dubourg, Vanderplas, Passos, Cournapeau, Brucher, Perrot and Duchesnay}]{scikitlearn}
\bibinfo{author}{Pedregosa, F.}, \bibinfo{author}{Varoquaux, G.}, \bibinfo{author}{Gramfort, A.}, \bibinfo{author}{Michel, V.}, \bibinfo{author}{Thirion, B.}, \bibinfo{author}{Grisel, O.}, \bibinfo{author}{Blondel, M.}, \bibinfo{author}{Prettenhofer, P.}, \bibinfo{author}{Weiss, R.}, \bibinfo{author}{Dubourg, V.}, \bibinfo{author}{Vanderplas, J.}, \bibinfo{author}{Passos, A.}, \bibinfo{author}{Cournapeau, D.}, \bibinfo{author}{Brucher, M.}, \bibinfo{author}{Perrot, M.}, \bibinfo{author}{Duchesnay, E.}, \bibinfo{year}{2011}.
\newblock \bibinfo{title}{Scikit-learn: Machine learning in {P}ython}.
\newblock \bibinfo{journal}{Journal of Machine Learning Research} \bibinfo{volume}{12}, \bibinfo{pages}{2825--2830}.
%Type = Article
\bibitem[{P{\'e}rez-Granados(2023)}]{perez2023birdnet}
\bibinfo{author}{P{\'e}rez-Granados, C.}, \bibinfo{year}{2023}.
\newblock \bibinfo{title}{Birdnet: applications, performance, pitfalls and future opportunities}.
\newblock \bibinfo{journal}{Ibis} \bibinfo{volume}{165}, \bibinfo{pages}{1068--1075}.
%Type = Article
\bibitem[{P{\'e}rez-Granados et~al.(2025)P{\'e}rez-Granados, Funosas, Morant, Mar{\'\i}n, Mendoza, Mohedano-Mu{\~n}oz, Santamar{\'\i}a, Bastianelli, M{\'a}rquez-Rodr{\'\i}guez, Budka et~al.}]{perez2025optimisation}
\bibinfo{author}{P{\'e}rez-Granados, C.}, \bibinfo{author}{Funosas, D.}, \bibinfo{author}{Morant, J.}, \bibinfo{author}{Mar{\'\i}n, O.H.}, \bibinfo{author}{Mendoza, I.}, \bibinfo{author}{Mohedano-Mu{\~n}oz, M.A.}, \bibinfo{author}{Santamar{\'\i}a, E.}, \bibinfo{author}{Bastianelli, G.}, \bibinfo{author}{M{\'a}rquez-Rodr{\'\i}guez, A.}, \bibinfo{author}{Budka, M.}, et~al., \bibinfo{year}{2025}.
\newblock \bibinfo{title}{Optimisation of passive acoustic bird surveys: a global assessment of birdnet settings} .
%Type = Inproceedings
\bibitem[{Piczak(2015a)}]{piczak2015esc}
\bibinfo{author}{Piczak, K.J.}, \bibinfo{year}{2015}a.
\newblock \bibinfo{title}{Esc: Dataset for environmental sound classification}, in: \bibinfo{booktitle}{Proceedings of the 23rd ACM international conference on Multimedia}, pp. \bibinfo{pages}{1015--1018}.
%Type = Inproceedings
\bibitem[{Piczak(2015b)}]{piczak2015dataset}
\bibinfo{author}{Piczak, K.J.}, \bibinfo{year}{2015}b.
\newblock \bibinfo{title}{{ESC}: {Dataset} for {Environmental Sound Classification}}, in: \bibinfo{booktitle}{Proceedings of the 23rd {Annual ACM Conference} on {Multimedia}}, \bibinfo{publisher}{{ACM Press}}. pp. \bibinfo{pages}{1015--1018}.
\newblock \URLprefix \url{http://dl.acm.org/citation.cfm?doid=2733373.2806390}, \DOIprefix\doi{10.1145/2733373.2806390}.
%Type = Article
\bibitem[{Rend{\'o}n et~al.(2008)Rend{\'o}n, Green, Aguilera and Almaraz}]{rendon2008status}
\bibinfo{author}{Rend{\'o}n, M.A.}, \bibinfo{author}{Green, A.J.}, \bibinfo{author}{Aguilera, E.}, \bibinfo{author}{Almaraz, P.}, \bibinfo{year}{2008}.
\newblock \bibinfo{title}{Status, distribution and long-term changes in the waterbird community wintering in do{\~n}ana, south--west spain}.
\newblock \bibinfo{journal}{Biological Conservation} \bibinfo{volume}{141}, \bibinfo{pages}{1371--1388}.
%Type = Article
\bibitem[{Rigoudy et~al.(2023)Rigoudy, Dussert, Benyoub, Besnard, Birck, Boyer, Bollet, Bunz, Caussimont, Chetouane et~al.}]{rigoudy2023deepfaune}
\bibinfo{author}{Rigoudy, N.}, \bibinfo{author}{Dussert, G.}, \bibinfo{author}{Benyoub, A.}, \bibinfo{author}{Besnard, A.}, \bibinfo{author}{Birck, C.}, \bibinfo{author}{Boyer, J.}, \bibinfo{author}{Bollet, Y.}, \bibinfo{author}{Bunz, Y.}, \bibinfo{author}{Caussimont, G.}, \bibinfo{author}{Chetouane, E.}, et~al., \bibinfo{year}{2023}.
\newblock \bibinfo{title}{The deepfaune initiative: a collaborative effort towards the automatic identification of european fauna in camera trap images}.
\newblock \bibinfo{journal}{European Journal of Wildlife Research} \bibinfo{volume}{69}, \bibinfo{pages}{113}.
%Type = Article
\bibitem[{Robbins(1981)}]{robbins1981effect}
\bibinfo{author}{Robbins, C.S.}, \bibinfo{year}{1981}.
\newblock \bibinfo{title}{Effect of time of day on bird activity}.
\newblock \bibinfo{journal}{Studies in avian biology} \bibinfo{volume}{6}, \bibinfo{pages}{275--286}.
%Type = Article
\bibitem[{Schuster et~al.(2024)Schuster, Walston and Little}]{schuster2024evaluation}
\bibinfo{author}{Schuster, G.E.}, \bibinfo{author}{Walston, L.J.}, \bibinfo{author}{Little, A.R.}, \bibinfo{year}{2024}.
\newblock \bibinfo{title}{Evaluation of an autonomous acoustic surveying technique for grassland bird communities in nebraska}.
\newblock \bibinfo{journal}{PloS one} \bibinfo{volume}{19}, \bibinfo{pages}{e0306580}.
%Type = Article
\bibitem[{Sethi et~al.(2024)Sethi, Bick, Chen, Crouzeilles, Hillier, Lawson, Lee, Liu, de~Freitas~Parruco, Rosten et~al.}]{sethi2024large}
\bibinfo{author}{Sethi, S.S.}, \bibinfo{author}{Bick, A.}, \bibinfo{author}{Chen, M.Y.}, \bibinfo{author}{Crouzeilles, R.}, \bibinfo{author}{Hillier, B.V.}, \bibinfo{author}{Lawson, J.}, \bibinfo{author}{Lee, C.Y.}, \bibinfo{author}{Liu, S.H.}, \bibinfo{author}{de~Freitas~Parruco, C.H.}, \bibinfo{author}{Rosten, C.M.}, et~al., \bibinfo{year}{2024}.
\newblock \bibinfo{title}{Large-scale avian vocalization detection delivers reliable global biodiversity insights}.
\newblock \bibinfo{journal}{Proceedings of the National Academy of Sciences} \bibinfo{volume}{121}, \bibinfo{pages}{e2315933121}.
%Type = Article
\bibitem[{Sossover et~al.(2024)Sossover, Burrows, Kahl and Wood}]{sossover2024using}
\bibinfo{author}{Sossover, D.}, \bibinfo{author}{Burrows, K.}, \bibinfo{author}{Kahl, S.}, \bibinfo{author}{Wood, C.M.}, \bibinfo{year}{2024}.
\newblock \bibinfo{title}{Using the birdnet algorithm to identify wolves, coyotes, and potentially their interactions in a large audio dataset}.
\newblock \bibinfo{journal}{Mammal Research} \bibinfo{volume}{69}, \bibinfo{pages}{159--165}.
%Type = Article
\bibitem[{Stowell(2022)}]{stowell2022computational}
\bibinfo{author}{Stowell, D.}, \bibinfo{year}{2022}.
\newblock \bibinfo{title}{Computational bioacoustics with deep learning: a review and roadmap}.
\newblock \bibinfo{journal}{PeerJ} \bibinfo{volume}{10}, \bibinfo{pages}{e13152}.
%Type = Article
\bibitem[{Stowell et~al.(2019)Stowell, Wood, Pamu{\l}a, Stylianou and Glotin}]{stowell2019automatic}
\bibinfo{author}{Stowell, D.}, \bibinfo{author}{Wood, M.D.}, \bibinfo{author}{Pamu{\l}a, H.}, \bibinfo{author}{Stylianou, Y.}, \bibinfo{author}{Glotin, H.}, \bibinfo{year}{2019}.
\newblock \bibinfo{title}{Automatic acoustic detection of birds through deep learning: the first bird audio detection challenge}.
\newblock \bibinfo{journal}{Methods in Ecology and Evolution} \bibinfo{volume}{10}, \bibinfo{pages}{368--380}.
%Type = Article
\bibitem[{Sugai et~al.(2019)Sugai, Silva, Ribeiro~Jr and Llusia}]{sugai2019terrestrial}
\bibinfo{author}{Sugai, L.S.M.}, \bibinfo{author}{Silva, T.S.F.}, \bibinfo{author}{Ribeiro~Jr, J.W.}, \bibinfo{author}{Llusia, D.}, \bibinfo{year}{2019}.
\newblock \bibinfo{title}{Terrestrial passive acoustic monitoring: review and perspectives}.
\newblock \bibinfo{journal}{BioScience} \bibinfo{volume}{69}, \bibinfo{pages}{15--25}.
%Type = Article
\bibitem[{Tokozume et~al.(2017)Tokozume, Ushiku and Harada}]{tokozume2017learning}
\bibinfo{author}{Tokozume, Y.}, \bibinfo{author}{Ushiku, Y.}, \bibinfo{author}{Harada, T.}, \bibinfo{year}{2017}.
\newblock \bibinfo{title}{Learning from between-class examples for deep sound recognition. arxiv 2017}.
\newblock \bibinfo{journal}{arXiv preprint arXiv:1711.10282} .
%Type = Article
\bibitem[{Tuia et~al.(2022)Tuia, Kellenberger, Beery, Costelloe, Zuffi, Risse, Mathis, Mathis, Van~Langevelde, Burghardt et~al.}]{tuia2022perspectives}
\bibinfo{author}{Tuia, D.}, \bibinfo{author}{Kellenberger, B.}, \bibinfo{author}{Beery, S.}, \bibinfo{author}{Costelloe, B.R.}, \bibinfo{author}{Zuffi, S.}, \bibinfo{author}{Risse, B.}, \bibinfo{author}{Mathis, A.}, \bibinfo{author}{Mathis, M.W.}, \bibinfo{author}{Van~Langevelde, F.}, \bibinfo{author}{Burghardt, T.}, et~al., \bibinfo{year}{2022}.
\newblock \bibinfo{title}{Perspectives in machine learning for wildlife conservation}.
\newblock \bibinfo{journal}{Nature communications} \bibinfo{volume}{13}, \bibinfo{pages}{1--15}.
%Type = Misc
\bibitem[{Ultralytics(2024)}]{ultralytics2024format}
\bibinfo{author}{Ultralytics}, \bibinfo{year}{2024}.
\newblock \bibinfo{title}{Ultralytics yolo format documentation}.
\newblock \bibinfo{howpublished}{\url{https://docs.ultralytics.com/datasets/detect/##ultralytics-yolo-format}}.
\newblock \bibinfo{note}{Accessed: 2025-01-27}.
%Type = Article
\bibitem[{Wieczorek et~al.(2012)Wieczorek, Bloom, Guralnick, Blum, D{\"o}ring, Giovanni, Robertson and Vieglais}]{wieczorek2012darwin}
\bibinfo{author}{Wieczorek, J.}, \bibinfo{author}{Bloom, D.}, \bibinfo{author}{Guralnick, R.}, \bibinfo{author}{Blum, S.}, \bibinfo{author}{D{\"o}ring, M.}, \bibinfo{author}{Giovanni, R.}, \bibinfo{author}{Robertson, T.}, \bibinfo{author}{Vieglais, D.}, \bibinfo{year}{2012}.
\newblock \bibinfo{title}{Darwin core: an evolving community-developed biodiversity data standard}.
\newblock \bibinfo{journal}{PloS one} \bibinfo{volume}{7}, \bibinfo{pages}{e29715}.
%Type = Article
\bibitem[{Wood and Kahl(2024)}]{wood2024guidelines}
\bibinfo{author}{Wood, C.M.}, \bibinfo{author}{Kahl, S.}, \bibinfo{year}{2024}.
\newblock \bibinfo{title}{Guidelines for appropriate use of birdnet scores and other detector outputs}.
\newblock \bibinfo{journal}{Journal of Ornithology} , \bibinfo{pages}{1--6}.
%Type = Article
\bibitem[{Wyse(2017)}]{wyse2017audio}
\bibinfo{author}{Wyse, L.}, \bibinfo{year}{2017}.
\newblock \bibinfo{title}{Audio spectrogram representations for processing with convolutional neural networks}.
\newblock \bibinfo{journal}{arXiv preprint arXiv:1706.09559} .
%Type = Article
\bibitem[{Xiao et~al.(2022)Xiao, Liu, Chen and Zhu}]{xiao2022amresnet}
\bibinfo{author}{Xiao, H.}, \bibinfo{author}{Liu, D.}, \bibinfo{author}{Chen, K.}, \bibinfo{author}{Zhu, M.}, \bibinfo{year}{2022}.
\newblock \bibinfo{title}{Amresnet: An automatic recognition model of bird sounds in real environment}.
\newblock \bibinfo{journal}{Applied Acoustics} \bibinfo{volume}{201}, \bibinfo{pages}{109121}.
%Type = Article
\bibitem[{Xie et~al.(2022)Xie, Zhao, Li, Ni and Zhang}]{xie2022kd}
\bibinfo{author}{Xie, J.}, \bibinfo{author}{Zhao, S.}, \bibinfo{author}{Li, X.}, \bibinfo{author}{Ni, D.}, \bibinfo{author}{Zhang, J.}, \bibinfo{year}{2022}.
\newblock \bibinfo{title}{Kd-cldnn: Lightweight automatic recognition model based on bird vocalization}.
\newblock \bibinfo{journal}{Applied Acoustics} \bibinfo{volume}{188}, \bibinfo{pages}{108550}.
%Type = Article
\bibitem[{Xie et~al.(2023)Xie, Zhong, Zhang, Liu, Ding and Triantafyllopoulos}]{xie2023review}
\bibinfo{author}{Xie, J.}, \bibinfo{author}{Zhong, Y.}, \bibinfo{author}{Zhang, J.}, \bibinfo{author}{Liu, S.}, \bibinfo{author}{Ding, C.}, \bibinfo{author}{Triantafyllopoulos, A.}, \bibinfo{year}{2023}.
\newblock \bibinfo{title}{A review of automatic recognition technology for bird vocalizations in the deep learning era}.
\newblock \bibinfo{journal}{Ecological Informatics} \bibinfo{volume}{73}, \bibinfo{pages}{101927}.
%Type = Article
\bibitem[{Zhang et~al.(2017)Zhang, Cisse, Dauphin and Lopez-Paz}]{zhang2017mixup}
\bibinfo{author}{Zhang, H.}, \bibinfo{author}{Cisse, M.}, \bibinfo{author}{Dauphin, Y.N.}, \bibinfo{author}{Lopez-Paz, D.}, \bibinfo{year}{2017}.
\newblock \bibinfo{title}{mixup: Beyond empirical risk minimization}.
\newblock \bibinfo{journal}{arXiv preprint arXiv:1710.09412} .

\end{thebibliography}

\section*{Declaration of generative AI and AI-assisted technologies in the writing process}

During the preparation of this work the authors used ChatGPT in order to enhance the writing quality and facilitate language translation. After using this tool/service, the authors reviewed and edited the content as needed and take full responsibility for the content of the publication.

\clearpage
\appendix
\noindent

\setcounter{figure}{0}
\setcounter{table}{0}

\section{Additional Technical Details}
\label{appendix:additional_info}

\subsection*{YOLOv8 Annotation Format and File Structure}

% Coordinates must be relative to the image dimensions and normalized. Each annotation was then converted into bounding boxes, where temporal coordinates were calculated based on the start and end times of the annotation, the duration of the audio segment (60 seconds) and the width of the spectrogram (930 pixels). For the frequency axis, the bounding box was centered along the y-axis of the spectrogram and spanned the entire height, represented as \textit{y\_center} = 0.5 and \textit{y\_height} = 1 in normalized terms. This process resulted in bounding box values represented as \textit{x\_center}, \textit{y\_center}, \textit{x\_width}, \textit{y\_height}, normalized between 0 and 1.  In line with YOLO’s official documentation \citep{ultralytics2024format}, annotations were stored in individual *.txt files for each spectrogram in a mirror directory corresponding to the audio files. The format of these files includes one row per annotation: \textit{class}, \textit{x\_center}, \textit{y\_center}, \textit{x\_width}, \textit{y\_height}. Each class is zero-indexed (starting from 0). Spectrograms without annotations have an empty *.txt file.

Annotations were created by manually labeling the temporal and frequency localization of vocalizations—i.e., using start and end times (in seconds) and frequency ranges (in Hz) relative to each 60-second audio file. These real-world values were then transformed into bounding boxes formatted according to the YOLOv8 object detection standard, where all coordinates are normalized with respect to the spectrogram image dimensions.

\begin{itemize}
    \item \textit{class}: the zero-indexed label corresponding to the annotated class,
    \item \textit{x\_center}, \textit{y\_center}: the normalized center coordinates of the bounding box,
    \item \textit{x\_width}, \textit{y\_height}: the normalized width and height of the bounding box.
\end{itemize}

Due to data limitations and to simplify the detection task, we chose to fix the vertical dimension of all bounding boxes to span the full frequency range of the spectrogram. This means each bounding box covers the entire vertical axis, and only the horizontal (temporal) dimension is variable. Thus, the vertical values were set as:

\[
y\_center = 0.5, \quad y\_height = 1.0
\]

Annotations were stored in individual \texttt{.txt} files, each corresponding to a spectrogram image. These files were placed in a mirrored directory structure matching the organization of the original audio files. In accordance with YOLOv8 standards \citep{ultralytics2024format}, each file contains one row per annotation in the following format:

\[
\textit{class}\quad x\_center\quad y\_center\quad x\_width\quad y\_height
\]

Spectrograms with no annotated vocalizations have an empty \texttt{.txt} file.

\subsection*{Temporal Conversion of Bird Song Detector Predictions}

Since the Bird Song Detector is implemented as a YOLOv8 object detector, its output consists of predicted bounding boxes in YOLOv8 format. To interpret these predictions in ecologically meaningful terms—i.e., as time intervals in the original audio—they must be converted back from normalized image coordinates to temporal coordinates in seconds.

Each spectrogram image corresponds to a 60-second audio segment and has fixed dimensions of \( W = 930 \) pixels in width (time axis) and \( H = 462 \) pixels in height (frequency axis). Therefore, the temporal resolution is approximately 0.0645 seconds per pixel.

YOLOv8 bounding boxes are returned in normalized format. Each box provides:
\begin{itemize}
    \item \( x_{\text{center}} \): the normalized center of the box along the time axis,
    \item \( w \): the normalized width of the box.
\end{itemize}

To compute the corresponding temporal boundaries in seconds, the following denormalization and conversion formulas are used:

\begin{equation*}
x_{\textit{center\_d}} = x_{\textit{center}} \times \textit{W}
\end{equation*}

\begin{equation*}
\textit{w\_d} = \textit{w} \times \textit{W}
\end{equation*}

\begin{equation*}
\textit{start\_sec} = \left( x_{\textit{center\_d}} - \frac{\textit{w\_d}}{2} \right) \times \frac{60}{\textit{W}}
\end{equation*}

\begin{equation*}
\textit{end\_sec} = \left( x_{\textit{center\_d}} + \frac{\textit{w\_d}}{2} \right) \times \frac{60}{\textit{W}}
\end{equation*}

Where \( W \) is the width of the spectrogram image used as input for the Bird Song Detector, set to \( W = 930 \) pixels in our case. The variables \( x_{\text{center}} \) and \( w \) are normalized values representing the bounding box center and width, respectively. To map these predictions to real-world time, they are first denormalized into \( x_{\textit{center\_d}} \) and \( w_{\textit{d}} \) by scaling with \( W \). The resulting values, \( \textit{start\_sec} \) and \( \textit{end\_sec} \), correspond to the start and end times (in seconds) of the detected bird song segment relative to that segment.

As shown in Figure \ref{fig:yoloconversion}, bounding boxes are first normalized for training and then later converted back into real-world time using the spectrogram width and duration. This visualization helps clarify the role of each variable in the forward and reverse transformation, and illustrates how the detector’s output aligns with the original audio.

\begin{figure*}[!ht]
    \centering
    \includegraphics[width=1\linewidth]{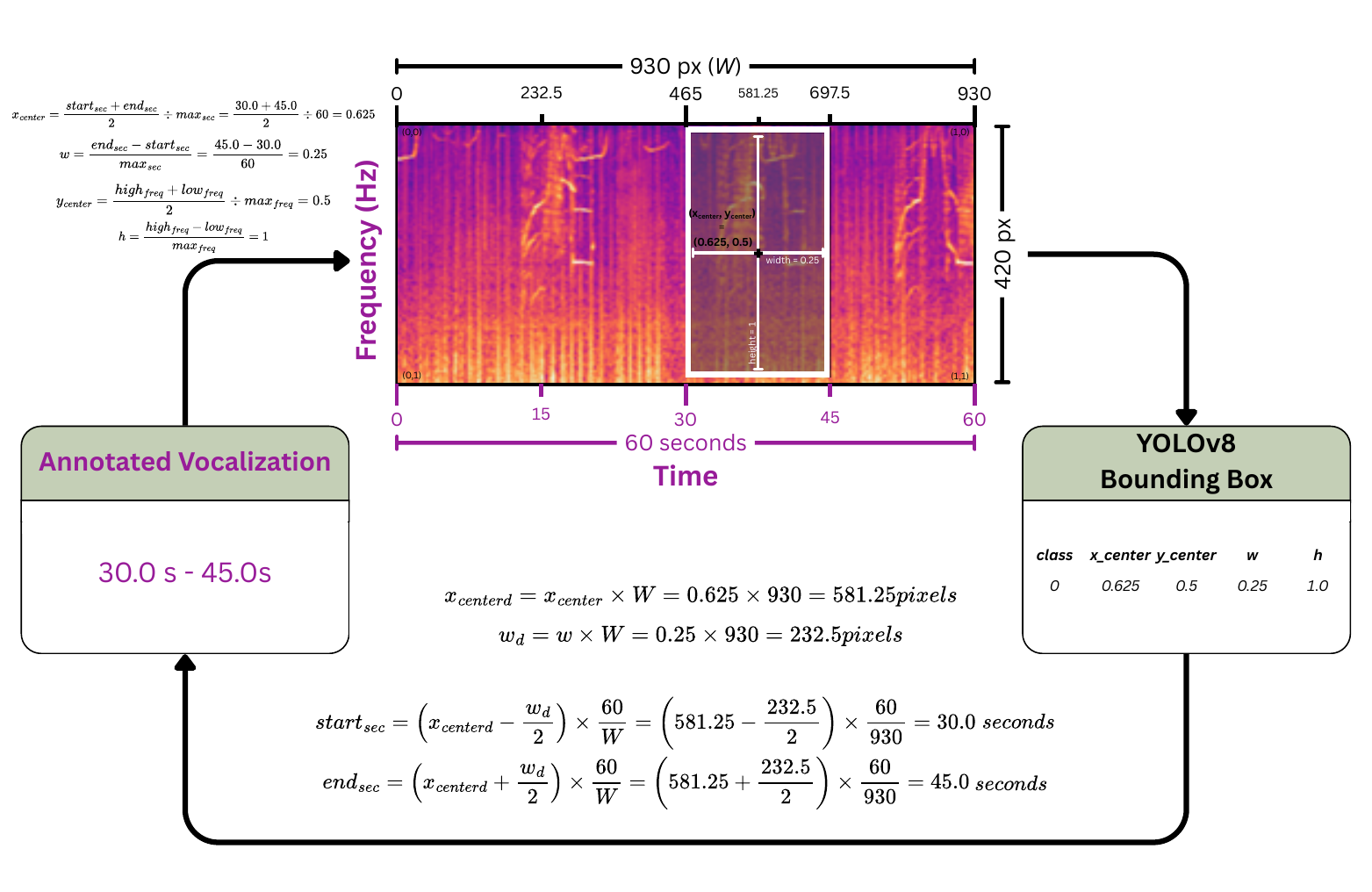}
    \caption{Visual representation of the YOLOv8 annotation format and its conversion between image-based bounding boxes and real-time bird vocalization intervals. The 60-second spectrogram shown includes an example annotation of a vocalization spanning from 30.0 to 35.0 seconds, corresponding to a width of 930 pixels (W). This annotated vocalization is normalized and converted to temporal coordinates based on YOLOv8 outputs, using the formulas displayed on the left side of the image. Conversely, YOLOv8 bounding boxes can be reconverted to temporal annotations using the formulas in the center of the image. Since all vocalizations are annotated across the full frequency range, the vertical axis of the bounding box remains constant across samples: \( y_{\text{center}} = 0.5 \), representing the normalized center of the image, and \( y_{\text{height}} = 1.0 \), spanning the full height of the image.}
    \label{fig:yoloconversion}
\end{figure*}

\clearpage

\setcounter{figure}{0}
\setcounter{table}{0}

\onecolumn

\section{Additional Figures and Tables}
\label{appendix:additional_data}

\FloatBarrier

\begin{figure}[!ht]
    \centering
    \includegraphics[width=1\linewidth]{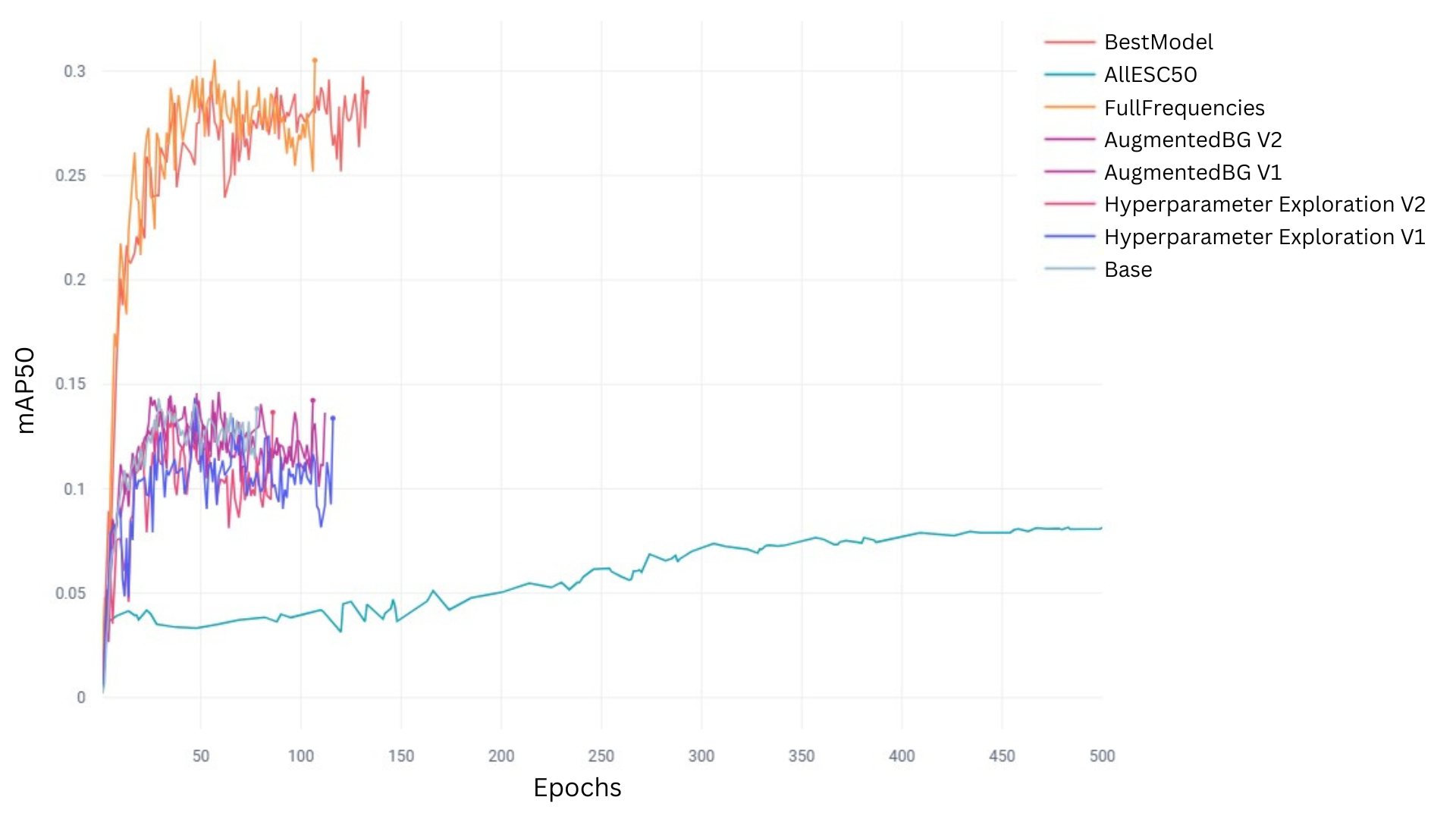}
    \caption{Mean Average Precision at 50\% Intersection over Union (mAP50) during training for different experiments of the Bird Song detector. Initial models (purple lines) showed high FP rates. Including a partial ESC50 dataset (dark orange line, \textit{Best Model}) significantly improved model performance. The worse values (green line) correspond to the whole inclusion of the ESC50 dataset.}
    \label{fig:training}
\end{figure}

\begin{table}[!ht]
\footnotesize
\centering
\begin{tabular*}{\textwidth}{@{\extracolsep{\fill}}|l|l|l|l|c|c|c|}
\hline
\multirow{2}{*}{\textbf{Name}} & \multicolumn{2}{c|}{\textbf{Background Augmentation}} & \multirow{2}{*}{\textbf{Frequency Spectrum}} & \multirow{2}{*}{\textbf{mAP50}} & \multirow{2}{*}{\textbf{Precision}} & \multirow{2}{*}{\textbf{Recall}} \\ \cline{2-3}
& \textbf{Synthetic Background Augmentation} & \textbf{ESC50 Dataset} & & & & \\ \hline
\textit{BestModel} & Add Noise + Intensity Change & Reduced & Full Spectrum & 0.29 & \textbf{0.412} & \textbf{0.308} \\ \hline
\textit{AllESC50} & Add Noise + Intensity Change & Full & Full Spectrum & 0.082 & 0.142 & 0.201 \\ \hline
\textit{FullFrequencies} & Add Noise + Intensity Change & \xmark & Full Spectrum & \textbf{0.305} & 0.399 & 0.302 \\ \hline
\textit{AugmentedBG V2} & Add Noise + Intensity Change & \xmark & Range Bounded & 0.142 & 0.291 & 0.163 \\ \hline
\textit{AugmentedBG V1} & Add Noise & \xmark & Range Bounded & 0.136 & 0.232 & 0.17 \\ \hline
\textit{Hyperparameter Exploration V2} & \xmark & \xmark & Range Bounded & 0.137 & 0.272 & 0.172 \\ \hline
\textit{Hyperparameter Exploration V1} & \xmark & \xmark & Range Bounded & 0.134 & 0.258 & 0.172 \\ \hline
\textit{Base} & \xmark & \xmark & Range Bounded & 0.138 & 0.275 & 0.174 \\ \hline
\end{tabular*}
\caption{Experimental results and configurations of the Bird Song Detector. The \xmark\ symbol indicates that no synthetic augmentation or dataset was applied in that experiment.}
\label{tab:experimental_results}
\end{table}

\begin{table}[!ht]
\footnotesize
\centering
\begin{tabularx}{0.7\textwidth}{|l|>{\centering\arraybackslash}X|>{\centering\arraybackslash}X|>{\centering\arraybackslash}X|>{\centering\arraybackslash}X|}
\hline
 & \textbf{Precision} & \textbf{Recall} & \textbf{F1-score} & \textbf{Number of test samples} \\ 
\hline
\textbf{\textit{Anthus pratensis}} & 0.16 & 1.00 & 0.28 & 42 \\ \hline
\textbf{\textit{Calandrella brachydactyla}} & 0.16 & 0.07 & 0.10 & 113 \\ \hline
\textbf{\textit{Carduelis carduelis}} & 0.00 & 0.00 & 0.00 & 1 \\ \hline
\textbf{\textit{Cettia cetti}} & 0.60 & 0.13 & 0.21 & 23 \\ \hline
\textbf{\textit{Chloris chloris}} & 1.00 & 0.08 & 0.15 & 12 \\ \hline
\textbf{\textit{Ciconia ciconia}} & 0.50 & 0.05 & 0.10 & 19 \\ \hline
\textbf{\textit{Cisticola juncidis}} & 0.17 & 0.14 & 0.15 & 7 \\ \hline
\textbf{\textit{Sylviidae}} & 0.33 & 0.06 & 0.10 & 17 \\ \hline
\textbf{\textit{Cyanopica cooki}} & 0.00 & 0.00 & 0.00 & 3 \\ \hline
\textbf{\textit{Emberiza calandra}} & 0.56 & 0.39 & 0.46 & 51 \\ \hline
\textbf{\textit{Falco tinnunculus}} & 0.00 & 0.00 & 0.00 & 2 \\ \hline
\textbf{\textit{Galerida theklae}} & 0.00 & 0.00 & 0.00 & 3 \\ \hline
\textbf{\textit{Galerida cristata}} & 0.70 & 0.47 & 0.56 & 30 \\ \hline
\textbf{\textit{Hippolais polyglotta}} & 0.00 & 0.00 & 0.00 & 4 \\ \hline
\textbf{\textit{Linaria cannabina}} & 0.00 & 0.00 & 0.00 & 1 \\ \hline
\textbf{\textit{Luscinia megarhynchos}} & 0.30 & 0.38 & 0.33 & 29 \\ \hline
\textbf{\textit{Melanocorypha calandra}} & 0.00 & 0.00 & 0.00 & 2 \\ \hline
\textbf{\textit{Merops apiaster}} & 0.00 & 0.00 & 0.00 & 3 \\ \hline
\textbf{\textit{Milvus migrans}} & 0.00 & 0.00 & 0.00 & 9 \\ \hline
\textbf{\textit{Motacilla flava}} & 0.00 & 0.00 & 0.00 & 2 \\ \hline
\textbf{\textit{Parus major}} & 0.00 & 0.00 & 0.00 & 2 \\ \hline
\textbf{\textit{Passer} sp.} & 0.00 & 0.00 & 0.00 & 8 \\ \hline
\textbf{\textit{Pica pica}} & 0.00 & 0.00 & 0.00 & 5 \\ \hline
\textbf{\textit{Saxicola rubicola}} & 0.47 & 0.30 & 0.37 & 23 \\ \hline
\textbf{\textit{Serinus serinus}} & 0.00 & 0.00 & 0.00 & 4 \\ \hline
\textbf{\textit{Streptopelia decaocto}} & 0.33 & 0.12 & 0.18 & 16 \\ \hline
\textbf{\textit{Sturnus} sp.} & 0.40 & 0.43 & 0.42 & 76 \\ \hline
\textbf{\textit{Turdus merula}} & 0.73 & 0.50 & 0.59 & 48 \\ \hline
\hline
\textbf{Accuracy} & & & \textbf{0.30} & 563 \\ \hline
\textbf{Macro Avg} & \textbf{0.21} & \textbf{0.14} & \textbf{0.13} & 563 \\ \hline
\textbf{Weighted Avg} & \textbf{0.37} & \textbf{0.30} & \textbf{0.28} & 563 \\ \hline
\end{tabularx}
\caption{Classification report for each class of the Bird Song Detector and a fine-tuned BirdNET approach, which was the best performing among all tested models. The table presents  precision, recall and F1-score for each class. The last three rows show overall performance metrics: Accuracy, Macro average, and Weighted average.}

\label{tab:classification_report}
\end{table}

\end{document}